\documentclass[aps,prd,twocolumn,groupedaddress,nofootinbib,longbibliography]{revtex4-1}

\usepackage{amssymb}
\usepackage{amsmath}
\usepackage[linktocpage=true]{hyperref}
\usepackage{graphicx}

\newcommand{\eqH}{\stackrel{\rm H}{=}}

\begin{document}

\title{Schwarzschild binary supported by an Appell ring}

\author{O. Semer\'ak}
\email[]{oldrich.semerak@mff.cuni.cz}

\author{M. Basovn\'{\i}k}
\email[]{mbasovnik@gmail.com}

\author{P. Kotla\v{r}\'{\i}k}
\email[]{kotlarik.petr@gmail.com}

\affiliation{Institute of Theoretical Physics, Faculty of Mathematics and Physics,
             Charles University, Prague, Czech Republic}

\date{\today}

\begin{abstract}
We continue to study black holes subjected to strong sources of gravity, again paying special attention to the behaviour of geometry in the black-hole interior. After examining, in previous two papers, the deformation arising in the Majumdar-Papapetrou binary of extremally charged black holes and that of a Schwarzschild black hole due to a surrounding (Bach-Weyl) ring, we consider here the system of two Schwarzschild-type black holes held apart by the Appell ring. After verifying that such a configuration can be in a strut-free equilibrium along certain lines in a parameter space, we compute several basic geometric characteristics of the equilibrium configurations. Then, like in previous papers, we calculate and visualize simple invariants determined by the metric (lapse or, equivalently, potential), by its first derivatives (gravitational acceleration) and by its second derivatives (Kretschmann scalar). Extension into the black-hole interior is achieved along particular null geodesics starting tangentially to the horizon. In contrast to the case involving the Bach-Weyl ring, here each single black hole is placed {\em asymmetrically} with respect to the equatorial plane (given by the Appell ring) and the interior geometry is really deformed in a non-symmetrical way. Inside the black holes, we again found regions of negative Kretschmann scalar in some cases.
\end{abstract}

\pacs{0420Jb, 0440Nr, 0470Bw}

\maketitle

\section{Introduction}

With the aim to study the geometry of black holes surrounded by other strong sources of gravity, we first \cite{SemerakB-16} considered the Majumdar-Papapetrou binary made of two extremely charged black holes and found that the horizon interior is not much deformed within this class of solutions. Therefore, in the second paper \cite{BasovnikS-16}, we tried to distort a black hole which is far from being extreme (actually the Schwarzschild one), surrounding it by a concentric static and axisymmetric thin ring (described by the aged Bach--Weyl solution). A much stronger effect was found inside such a black hole, including an occurrence of the regions of negative Kretschmann invariant.

Keeping the static and axially symmetric setting, we turn to another Weyl-type superposition in the present paper, namely a binary of Schwarzschild black holes held apart by an Appell ring. As opposed to the previous paper treating the black hole surrounded by a ring, the present system is hardly relevant astrophysically, but very interesting theoretically, because it provides a strut-free possibility of keeping two uncharged (and non-rotating) black holes in static equilibrium. The cost is the inclusion of a source (the Appell ring) which produces, similarly like the ({\em rotating}) Kerr source, a ``repulsive" field in the central part, and whose interpretation rests on either a negative-mass density layer or a double-sheet topology.

Below, we first give the metric functions for the Schwarzschild and Appell space-times in appropriate coordinates, and describe how to make their non-linear superposition corresponding to the configuration described above (section \ref{metric-functions}). In section \ref{equilibrium-configurations}, we show that such a configuration can be in a strut-free equilibrium, find the lines of these equilibria in a parameter space and compute their basic geometric measures. Basic properties of the horizons are given in section \ref{horizons}. After briefly exemplifying, in section \ref{BH-exterior}, the shape of the geometry outside the black holes, we show, in section \ref{BH-interior}, how the metric can be extended to also describe the interior of (any of) the black holes and illustrate how the geometry is affected there by the other sources. In particular, we compute the free-fall time from the opposite poles of the horizon to the singularity, and plot the basic invariants determined by the metric (lapse $N$) and by its first and second derivatives (gravitational-acceleration scalar $\kappa^2$ and Kretschmann scalar $K$).

The (vacuum) static and axisymmetric problem is most easily solvable in the Weyl coordinates $(t,\rho,z,\phi)$, $t$ and $\phi$ being Killing time and azimuth, and $\rho$ and $z$ covering, isotropically, the meridional planes (orthogonal to both Killing directions). In such coordinates, the metric reads
\[{\rm d}s^2=-N^2{\rm d}t^2+\frac{\rho^2}{N^2}\,{\rm d}\phi^2
             +\frac{e^{2\lambda}}{N^2}\left({\rm d}\rho^2+{\rm d}z^2\right)\]
and only contains two unknown functions $N\!\equiv\!e^\nu$ and $\lambda$ of $\rho$ and $z$, of which $\nu$, representing the Newtonian potential, is given by solution of Laplace equation, so it superposes linearly when the field of multiple sources is being seeked.\footnote
{We do not repeat the whole basics on Weyl space-times, please see for example the first paper \cite{SemerakB-16} of this series, including references.}
For the Weyl-type metric, the above invariants $N$, $\kappa$ and $K$ are given by
\begin{align}
  &N^2\equiv e^{2\nu}:=-g_{tt} \,, \\
  &\kappa^2:=g^{\mu\nu}N_{,\mu}N_{,\nu}
            =\frac{N^2}{e^{2\lambda}}\left[(N_{,\rho})^2+(N_{,z})^2\right],
            \label{acceleration} \\
  &\frac{e^{4\lambda-4\nu}}{16}\,K
      :=\frac{e^{4\lambda-4\nu}}{16}\;
        R_{\mu\nu\kappa\lambda}R^{\mu\nu\kappa\lambda}=  \nonumber \\
      &=(\nu_{,\rho\rho})^2+(\nu_{,zz})^2+(\nu_{,\rho z})^2
          +\nu_{,\rho\rho}\nu_{,zz} \nonumber \\
      &\quad +3(1\!-\!\rho\nu_{,\rho})\left[(\nu_{,\rho})^2\!+\!(\nu_{,z})^2\right]^2
          \!+\!\rho^2\left[(\nu_{,\rho})^2\!+\!(\nu_{,z})^2\right]^3 \nonumber \\
      &\quad +3\nu_{,\rho\rho}(\nu_{,\rho})^2+3\nu_{,zz}(\nu_{,z})^2
          +6\nu_{,\rho z}\nu_{,\rho}\nu_{,z} \nonumber \\
      &\quad +\rho\nu_{,\rho}\left[3(\nu_{,z})^2-(\nu_{,\rho})^2\right]
           (\nu_{,\rho\rho}-\nu_{,zz}) \nonumber \\
      &\quad +2\rho\nu_{,\rho z}\nu_{,z}\left[(\nu_{,z})^2-3(\nu_{,\rho})^2\right].
      \label{Kretschmann-static-Weyl}
\end{align}

\section{Weyl-metric functions for Schwarzschild and for the Appell ring}
\label{metric-functions}

Following e.g. \cite{SemerakZZ-99a}, let us remind the Weyl-coordinate form of the Schwarzschild and Appell-ring metrics.
The first of them is given by
\begin{align}
  \nu_{\rm Schw} &= \frac{1}{2}\,\ln\frac{d_1+d_2-2M}{d_1+d_2+2M} \\
                 &= \frac{1}{2}\,\ln\left(1-\frac{2M}{r}\right),  \label{nuSchw} \\
  \lambda_{\rm Schw} &= \frac{1}{2}\,\ln\frac{(d_1+d_2)^2-4M^2}{4\,d_1 d_2} \\
                     &= \frac{1}{2}\,\ln\frac{r(r-2M)}{r(r-2M)+M^2\sin^2\theta} \; ,
                        \label{lambdaSchw}
\end{align}
where $M$ is the mass parameter,
\[d_{1,2}:=\sqrt{\rho^2+(z\mp M)^2}=r-M\mp M\cos\theta \,,\]
and the second expressions are in Schwarzschild coordinates $(r,\theta)$.
The Schwarzschild $\rightarrow$ Weyl transformation reads
\begin{equation}  \label{Schw-Weyl}
  \rho=\sqrt{r(r-2M)}\,\sin\theta, \qquad
  z=(r-M)\,\cos\theta
\end{equation}
and its inverse, above the horizon,
\begin{equation}  \label{Weyl-Schw}
  r-M=\frac{d_2+d_1}{2} \;, \qquad
  M\cos\theta=\frac{d_2-d_1}{2} \;.
\end{equation}

The simplest Appell-ring metric is given by
\begin{align}
 \nu_{\rm App} &= -\frac{\cal M}{\sqrt{2}\,\Sigma}\,\sqrt{\Sigma+\rho^2+z^2-a^2} \\
               &= -\frac{{\cal M}R}{\Sigma} \;,  \label{nuApp} \\
 \lambda_{\rm App}
    &= \frac{{\cal M}^2}{8a^2}
       \left[1\!-\frac{\rho^2\!+\!z^2\!+\!a^2}{\Sigma}
              -\frac{2a^2\rho^2(\Sigma^2\!-\!8z^2 a^2)}{\Sigma^4}\right] \\
    &= -\frac{{\cal M}^2\sin^2\vartheta}{4\,\Sigma}\!
       \left[1\!+\frac{(R^2\!+\!a^2)(\Sigma^2\!-\!8R^2 a^2\cos^2\vartheta)}{\Sigma^3}\right]\!,
       \label{lambdaApp}
\end{align}
where ${\cal M}$ is the mass and $a$ is the Weyl radius of the ring,
\begin{align*}
  &l_{1,2}:=\sqrt{(\rho\mp a)^2+z^2}=\sqrt{R^2+a^2}\mp a\sin\vartheta \,,\\
  &\Sigma :=l_1 l_2
           =\sqrt{(\rho^2-a^2+z^2)^2+4a^2 z^2}
           =R^2+a^2\cos^2\vartheta \,,
\end{align*}
and the second expressions are in ellipsoidal (oblate spheroidal) coordinates $(R,\vartheta)$.
The oblate $\rightarrow$ Weyl transformation reads
\begin{equation}  \label{oblate}
  \rho=\sqrt{R^2+a^2}\,\sin\vartheta, \qquad
  z=R\cos\vartheta,
\end{equation}
and its inverse
\begin{equation}
  \sqrt{R^2+a^2}=\frac{l_2+l_1}{2} \;, \qquad
  a\,\sin\vartheta=\frac{l_2-l_1}{2} \;.
\end{equation}
General relativistic space-times of Appell rings (this type of solution appeared in electrostatics originally) have been mainly treated by \cite{GleiserP-89} (see also \cite{Semerak-16}). As discussed and illustrated in \cite{SemerakZZ-99a} (see Appendix A there), the spatial structure of the Appell solution is similar to that of the Kerr solution (where, however, $\rho$ and $z$ must be understood as the Kerr-Schild cylindrical coordinates rather than the Weyl ones), but there is (of course) no horizon and no rotational dragging. In particular, both space-times have the disc $(z\!=\!0,\rho\!\leq\!a)$ $\Leftrightarrow$ $R\!=\!0$ at their centre, which is intrinsically flat but whose ring-like boundary $(z\!=\!0,\rho\!=\!a)$ $\Leftrightarrow$ $(R\!=\!0,\vartheta\!=\!\pi/2)$ represents a curvature singularity ($\Sigma\!=\!0$). If approaching the disc from either side (along $\vartheta\neq\pi/2$), $R$ decreases to zero, whereas its gradient does not vanish, which has to be interpreted either as a presence of a layer of mass with effective surface density
\begin{equation}
  \sigma =-\frac{{\cal M}a}{2\pi(a^2-\rho^2)^{3/2}}
         =-\frac{\cal M}{2\pi a^2\cos^3\vartheta} \;,
\end{equation}
or as an indication that the manifold continues, across the disc serving as a branch cut, smoothly to the second asymptotically flat sheet characterized by $R\!<\!0$. The double-sheeted topology may seem artificial, but only before one realizes that the above density is everywhere {\em negative} and diverges to $-\infty$ toward the disc edge; at the very singular rim it finally jumps to $+\infty$, to yield the finite positive total mass ${\cal M}$. Irrespectively of the interpretation, in the spherical region $\rho^2+z^2\!<\!a^2$ $\Leftrightarrow$ $0\!\leq\!R\!<\!a|\!\cos\vartheta|$ the field is ``repulsive" in the sense that momentarily static particles (those at rest with respect to infinity) are accelerated {\em away} from the central disc.

It is obvious from (\ref{nuSchw}) and (\ref{nuApp}) that both $\nu_{\rm Schw}$ and $\nu_{\rm App}$ are everywhere negative, just with $\nu_{\rm App}\!=\!0$ in the interior of the ring (on $z\!=\!0$, $\rho\!<\!a$).
The second function $\lambda$ is negative for the Schwarzschild field alone (\ref{lambdaSchw}), while for the pure Appell field it is also positive in a certain region. Actually, for $a^2\cos^2\vartheta\!=\!3R^2$ (thus $\Sigma\!=\!4R^2$), for example, (\ref{lambdaApp}) yields $\lambda\!=\!\frac{{\cal M}^2\sin^2\vartheta}{128\,R^4}\,(a^2-7R^2)$, which for a sufficiently small $R$ is positive. Along the symmetry axis ($\rho\!=\!0$), both $\lambda_{\rm Schw}$ and $\lambda_{\rm App}$ vanish, except for the divergence of $\lambda_{\rm Schw}$ along the black-hole segments; in the Appell-ring plane ($z\!=\!0$), $\lambda_{\rm App}$ is otherwise everywhere negative, both inside and outside the ring.
Basic properties of the Appell ring has been visualized in \cite{Semerak-16} (in comparison with several other thin-ring solutions).

\begin{figure}
\centering
\includegraphics[width=\columnwidth]{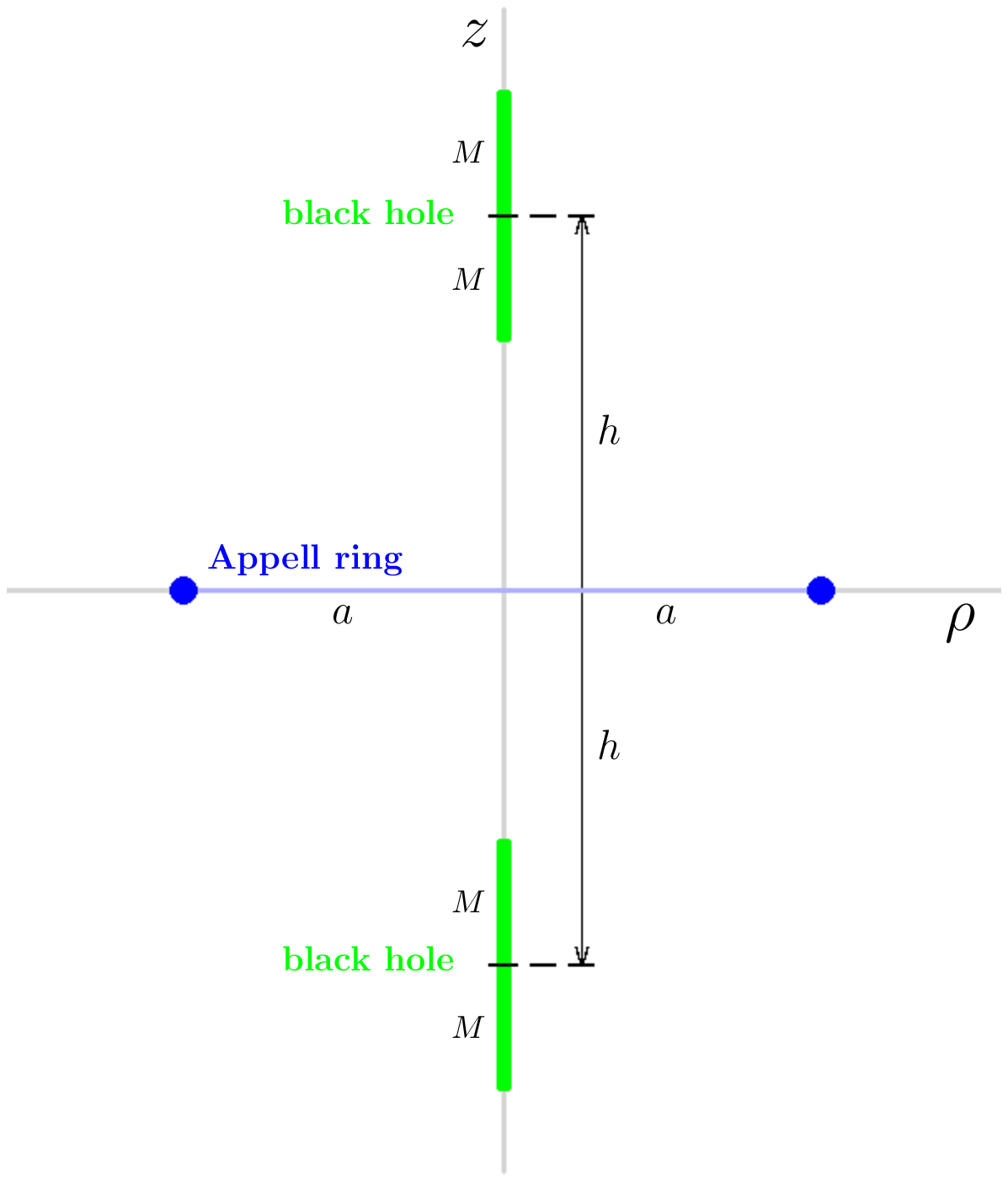}
\caption
{A Weyl-coordinate scheme of the configuration considered (its meridional section), made of two equal black holes placed symmetrically with respect to an Appell ring. Relevant dimensions are labelled.}
\label{scheme}
\end{figure}

\subsection{Superposition}

In the present paper, we first want to check the intuition that the central repulsive region of the Appell space-time could serve as a ``support" for two (attractive) sources (see figure \ref{scheme}). Indeed, for {\em test} particles, the locations $\rho\!=\!0$, $z\!=\!\pm a$, symmetrically placed with respect to the Appell centre, are {\em equilibrium} (the particles left there with zero velocity will stay there). In the Weyl coordinates, the Appell source is a ring $(z\!=\!0,\rho\!=\!a)$, while the Schwarzschild black hole is a massive line segment $(\rho\!=\!0,|z|\!\leq\!M)$. Rather than to estimate that $(\rho\!=\!0,z\!=\!\pm a)$ will remain equilibrium locations also for {\em heavy} bodies (actually black holes in our case), let us consider a generic location $(\rho\!=\!0,z\!=\!\pm h)$. Placing two (same) Schwarzschild holes there requires to shift them to $z\rightarrow z\pm h$ (with $\rho$ kept zero) and then superpose like
\begin{equation}  \label{superposition}
  \nu = \nu_{\rm App}(z)+\nu_{\rm Schw}(z\!-\!h)+\nu_{\rm Schw}(z\!+\!h)
  \quad {\rm with} \;\; M<h \,.
\end{equation}
Being given by very simple functions, the resulting potential is easily verified to have a reasonable and expectable shape, with divergences only found at the sources themselves. Clearly $\nu\!<\!0$ everywhere.

However, $\nu$ is just the first, Newtonian part of the story. The relativistic field is also described by the second metric function $\lambda$ which does {\em not} add up linearly and which can also give rise to various features including singularities. It is given by line integrals of
\begin{equation}  \label{lambda-eqs}
  \lambda_{,\rho}=\rho\,(\nu_{,\rho})^2-\rho\,(\nu_{,z})^2 \,, \qquad
  \lambda_{,z}=2\rho\,\nu_{,\rho}\nu_{,z} \,,
\end{equation}
which are usually tackled starting from some vacuum part of the symmetry axis where $\lambda$ has to vanish.
At several special locations, one finds
\begin{align*}
  &\lambda_{,z}\sim z      &{\rm at}& \; z=0, \;\, \rho>a, \\
  &\lambda_{,z}\sim \rho^2 &{\rm at}& \; \rho=0,
                                      \; \mbox{{\rm except~along~black-hole~parts}}, \\
  &\lambda_{,z}=2\nu_{,z}  &{\rm at}& \; \rho=0, \; h\!-\!M\leq|z|\leq h\!+\!M \;(\rm black~holes), \\
  &\lambda_{,\rho}\sim\rho &{\rm at}& \; \rho=0,
                                      \; \mbox{{\rm except~along~black-hole~parts}}, \\
  &\lambda_{,\rho}\simeq\rho^{-1} &{\rm at}& \; \rho=0, \; h\!-\!M\leq|z|\leq h\!+\!M \;(\rm black~holes).
\end{align*}

On the central circle $z\!=\!0$, $0\!<\!\rho\!<\!a$, the gradient of $\lambda$ is obtained from (\ref{lambda-eqs}) by using
\begin{widetext}
\begin{align}
  &\nu_{,\rho}=\frac{2M\rho\,\left[\sqrt{\rho^2+(h-M)^2}+\sqrt{\rho^2+(h+M)^2}\right]}
                    {(\rho^2+h^2-M^2)^2+4M^2\rho^2+(\rho^2+h^2-M^2)\sqrt{(\rho^2+h^2-M^2)^2+4M^2\rho^2}} \;, \\
  &\nu_{,z}(z\!\rightarrow\!0^\pm)=\mp\frac{{\cal M}\,a}{(a^2-\rho^2)^{3/2}} \;,
\end{align}
which can be integrated to get
\begin{equation}
  e^\lambda
  = e^{\lambda(\rho=0,z=0)}
    \left[\frac{1}{2}+\frac{\rho^2(h^2+M^2)+(h^2-M^2)^2}
                           {2\,(h^2-M^2)\,\sqrt{(\rho^2+h^2-M^2)^2+4M^2\rho^2}}\right]
    \exp\left[-\frac{{\cal M}^2\rho^2\,(2a^2-\rho^2)}{4a^2\,(a^2-\rho^2)^2}\right].
  \label{explambda,circle}
\end{equation}
\end{widetext}
After multiplying this by
\begin{equation}
  e^{-\nu}=\frac{\sqrt{\rho^2+(h-M)^2}+\sqrt{\rho^2+(h+M)^2}+2M}
                {\sqrt{\rho^2+(h-M)^2}+\sqrt{\rho^2+(h+M)^2}-2M}
\end{equation}
(remember we are still on the circle $z\!=\!0$, $0\!<\!\rho\!<\!a$),
one can ask about proper radius of the central ring, $\int_0^a e^{\lambda-\nu}(z\!=\!0)\,{\rm d}\rho$.
The integrand is everywhere finite on the central circle, starting from
\[e^{\lambda-\nu}(\rho\!=\!0,z\!=\!0)=e^\lambda(\rho\!=\!0,z\!=\!0)\,\frac{h+M}{h-M}\]
and falling to zero quickly towards $\rho\!\rightarrow\!a^-$, so the proper radius is {\em finite}.

The second simple measure is the Appell-ring proper circumference, as taken from its inside ($z\!=\!0$, $\rho\!\rightarrow\!a^-$), which is given by
\begin{align}
  &\int\limits_0^{2\pi}\sqrt{g_{\phi\phi}(z\!=\!0,\rho\!\rightarrow\!a^-)}\;{\rm d}\phi
   =\int\limits_0^{2\pi}\rho\,e^{-\nu(z=0,\rho\rightarrow a^-)}\;{\rm d}\phi=  \nonumber \\
  &=2\pi a\,e^{-\nu(z=0,\rho\rightarrow a^-)}= \nonumber \\
  &=2\pi a\;\frac{\sqrt{a^2+(h-M)^2}+\sqrt{a^2+(h+M)^2}+2M}{\sqrt{a^2+(h-M)^2}+\sqrt{a^2+(h+M)^2}-2M} \;.
\end{align}
Clearly this is finite, in contrast to the circumference computed from outside ($z\!=\!0$, $\rho\!\rightarrow\!a^+$) which is infinite. In the next section, we illustrate how the ring's circumference and radius depend on parameters for {\em equilibrium} (strut-free) configurations.

Outside of the ring, at $z\!=\!0$, $\rho\!>\!a$, one has $\nu_{,z}\!=\!0$, hence $\lambda_{,\rho}\!=\!\rho\,(\nu_{,\rho})^2$. This is slightly longer explicitely, but anyway it behaves like $\lambda_{,\rho}\sim(\rho-a)^{-3}$ at $\rho\!\rightarrow\!a^+$, which means that $\lambda(z\!=\!0)\sim-(\rho-a)^{-2}$ in this limit. Since, on the other hand, $\nu(z\!=\!0)\sim-(\rho-a)^{-1/2}$ there, the proper radial distance to the ring measured from outside along the equatorial plane, $\int_a^{\rho>a}e^{\lambda-\nu}(z\!=\!0)\,{\rm d}\rho$, is also {\em finite}. Due to the strong exponential damping brought by $\lambda$, the proper equatorial radius almost does not change in the ring's vicinity (this holds from both sides). This is very different from the other aged ring solution, the Bach--Weyl ring (we considered it previously in \cite{BasovnikS-16,Semerak-16}), which is at finite proper distance from outside, but infinitely far when approached from below. Let us add that the $\nu(z\!=\!0)\sim-(\rho-a)^{-1/2}$ behaviour also means that the proper circumference of the ring $2\pi a\,e^{-\nu}(z\!=\!0)$ is infinite if taken from outside $(\rho\!\rightarrow\!a^+)$, as mentioned already. (On the other hand, the circumference of the Bach--Weyl ring is infinite, whether taken from inside or outside -- see \cite{Semerak-16}.)

A crucial task in static superpositions is to check whether there are no supporting ``struts" whose presence indicates that the given system is artificial and could not by itself remain static. In the axially symmetric case, such a check naturally starts on the symmetry axis (see e.g. \cite{LetelierO-98}). The axis is regular if the symmetric sections $z\!=\!0$ are flat in its neighbourhood, which requires $\lambda$ to vanish at $\rho\!=\!0$.

\section{Equilibrium configurations}
\label{equilibrium-configurations}

\begin{figure}
\includegraphics[width=\columnwidth]{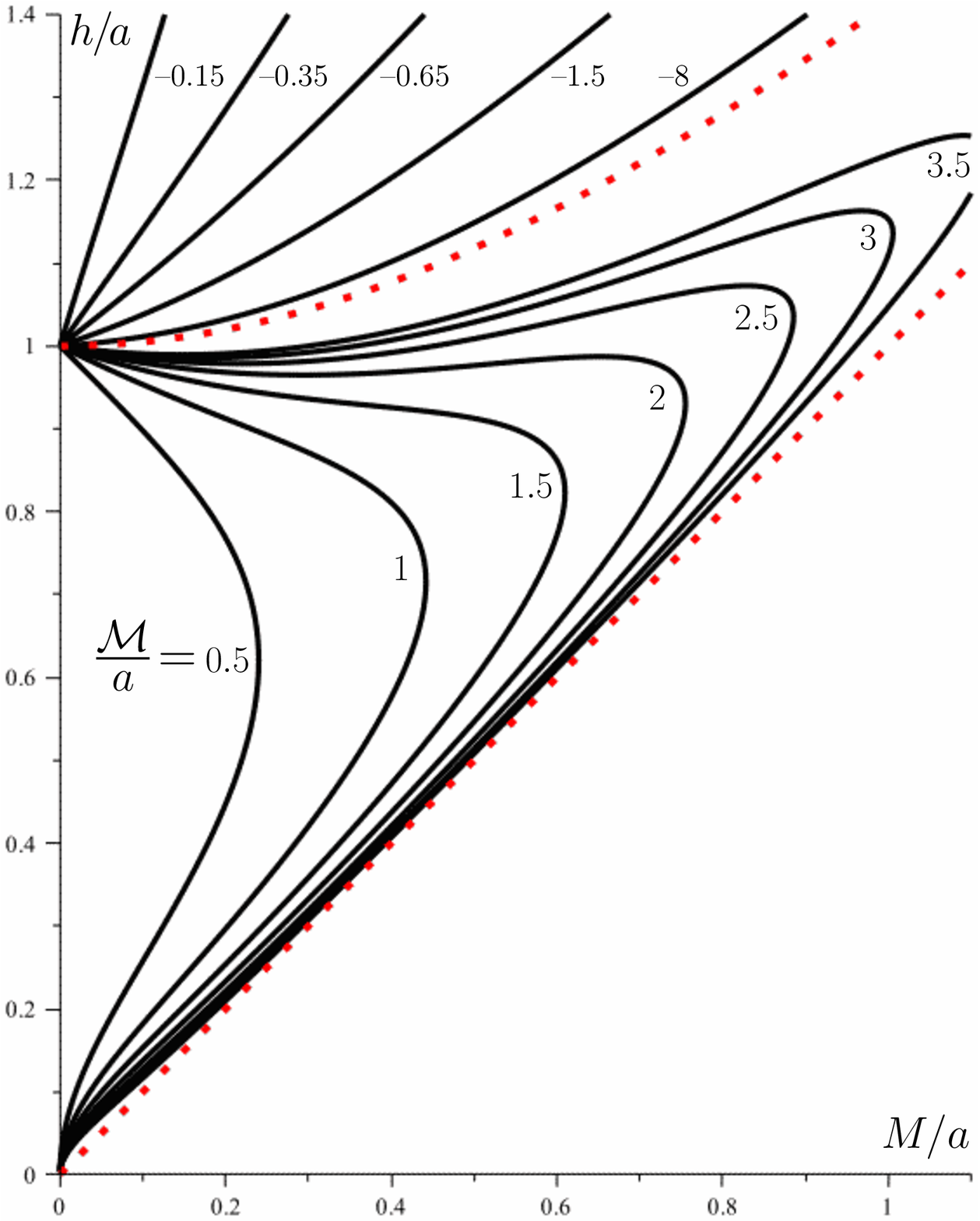}
\caption
{Lines of the strut-free equilibria within the $(M,h)$ plane, where $z\!=\!\pm h$ is the position of the black holes on the symmetry axis and $M$ is their mass, both given in units of the ring radius $a$. Curves obtained for several different ring masses ${\cal M}/a$ are given. Below the dotted red line $h\!=\!M$ the equilibrium is not possible, because the black holes would be too close, reaching down to $z\!=\!0$ or even beyond. The red dotted hyperbola $h^2\!=\!M^2+a^2$ divides the cases ${\cal M}\!>\!0$ (below) and ${\cal M}\!<\!0$ (above). In the ${\cal M}\!>\!0$ case, in the configurations situated to the left of the curves (``inside" them) the black holes would be repelled, while in those situated to the right of the curves (``outside" them) the holes would fall towards each other. Consequently, the larger-$h$ branches of equilibria are stable whereas the lower-$h$ equilibria are unstable in the vertical direction.}
\label{equilibria}
\end{figure}

\begin{figure}
\includegraphics[width=\columnwidth]{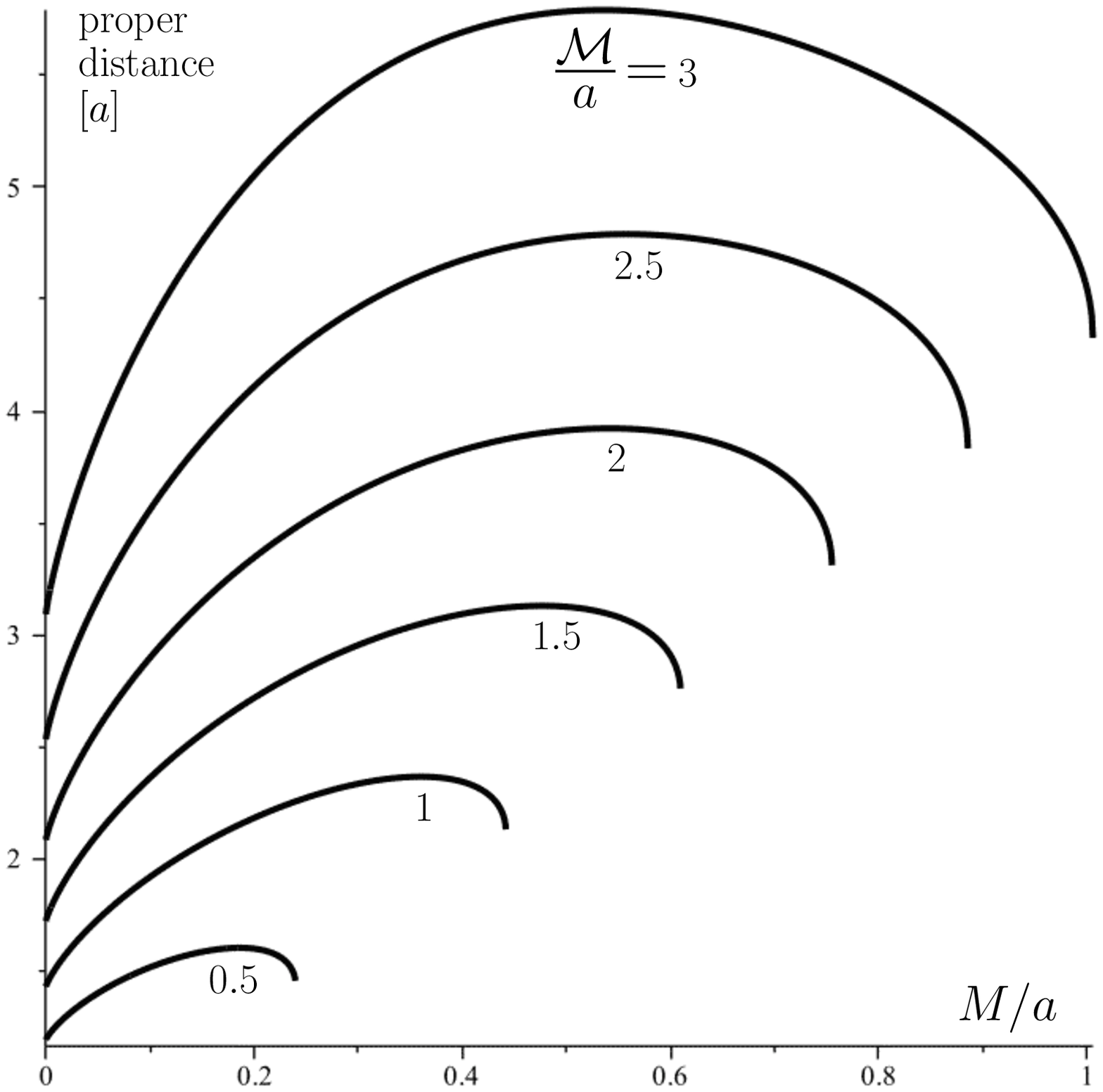}
\caption
{Proper distance from the Appell-ring centre $(\rho\!=\!0,z\!=\!0)$ to the black-hole horizons, drawn in dependence on the black-hole mass $M$ for several values of the Appell-ring mass ${\cal M}$, for the equilibrium configurations shown in figure \ref{equilibria} (we already consider just the vertically stable branches of the ${\cal M}\!>\!0$ case).}
\label{BH-distance}
\end{figure}

\begin{figure}
\includegraphics[width=\columnwidth]{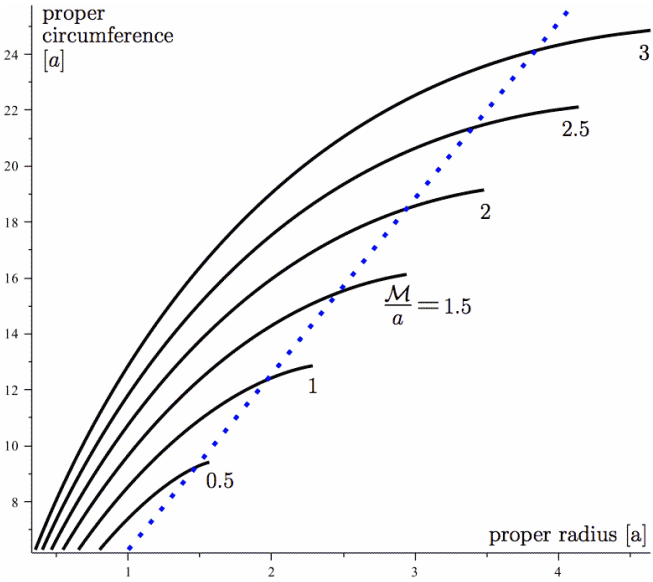}
\caption
{Proper interior circumference ($z\!=\!0$, $\rho\!\rightarrow\!a^-$) of the central Appell ring, drawn against the ring's proper radius. Both the radius and the circumference grow with black-hole mass $M$ which parametrizes the curves. The dotted blue line indicates Euclidean relation given by $2\pi r$; in Weyl coordinates, the radius is kept unit ($\rho\!=\!a$) and the circumference remains $2\pi a$, which would correspond just to the starting point of the dotted line. We again consider the equilibrium configurations shown in figure \ref{equilibria}, specifically, the vertically stable branches of the ${\cal M}\!>\!0$ case.}
\label{radius-circumference}
\end{figure}

\begin{figure*}
\includegraphics[width=\textwidth]{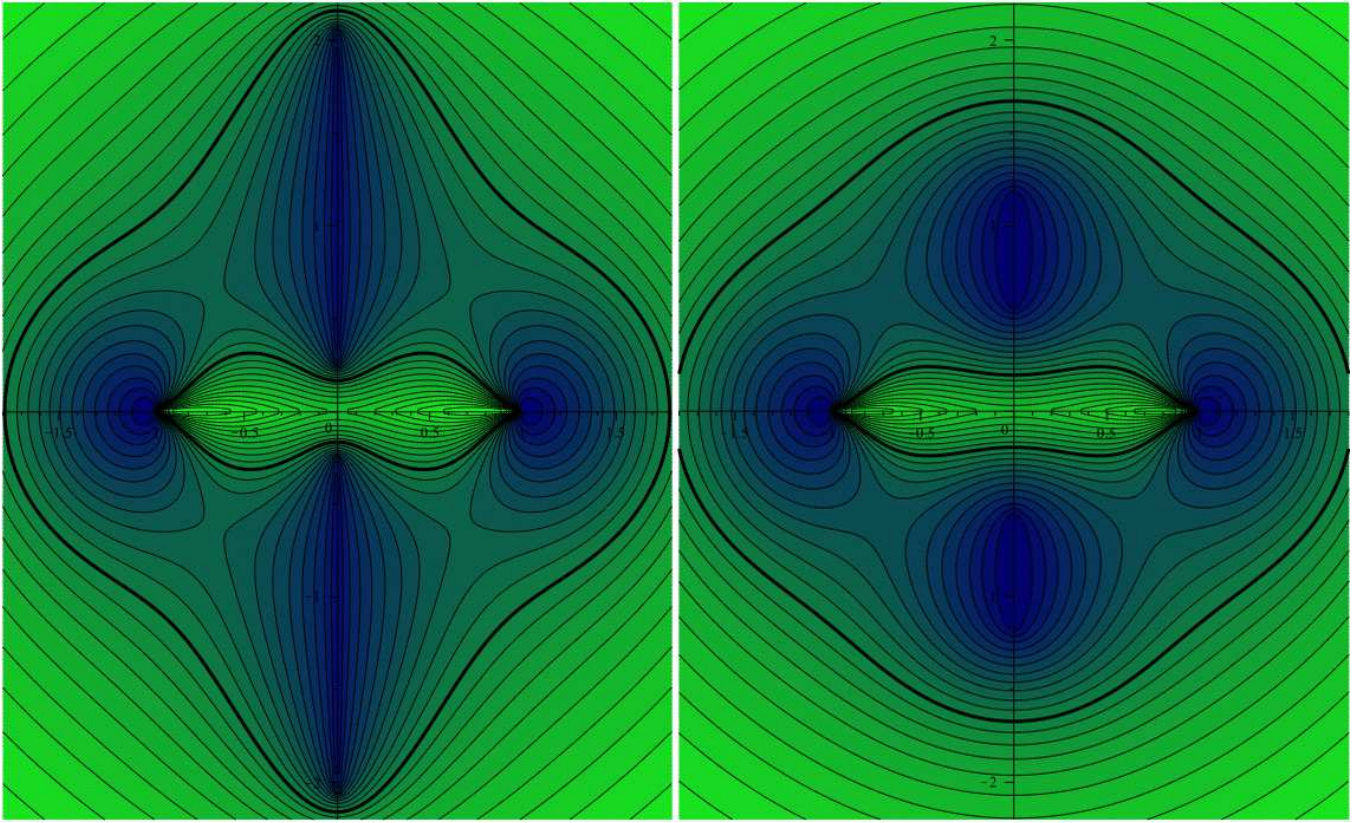}
\caption
{Meridional picture in the Weyl coordinates $\rho$, $z$ of the equipotential contours $\nu\!=\!{\rm const}$ for one heavier and one lighter realization of the equilibrium of the Appell ring and two symmetric black holes. In the left plot, ${\cal M}\!=\!3a$, $M\!=\!0.9a$ and $h\!\doteq\!1.1515a$, while in the right plot, ${\cal M}\!=\!a$, $M\!=\!0.3a$ and $h\!\doteq\!0.8713a$. Equatorial plane (given by the ring) is horizontal ($\rho$) and symmetry axis vertical ($z$), with the length unit given by the ring radius $a$ which is clearly visible. The potential values drawn has been chosen ``by hand"; they range from $-7.50$ to $-1.75$ in the left plot, while within three times weaker values (from $-2.50$ to $-0.58$) in the right plot. Deep potential is dark blue, while more shallow one is light green. For easier orientation, we have emphasized $\nu\!=\!-2.86$ contour in the left plot and the corresponding $\nu\!=\!-0.95$ contour in the right plot.}
\label{potential}
\end{figure*}

\begin{figure*}
\includegraphics[width=\textwidth]{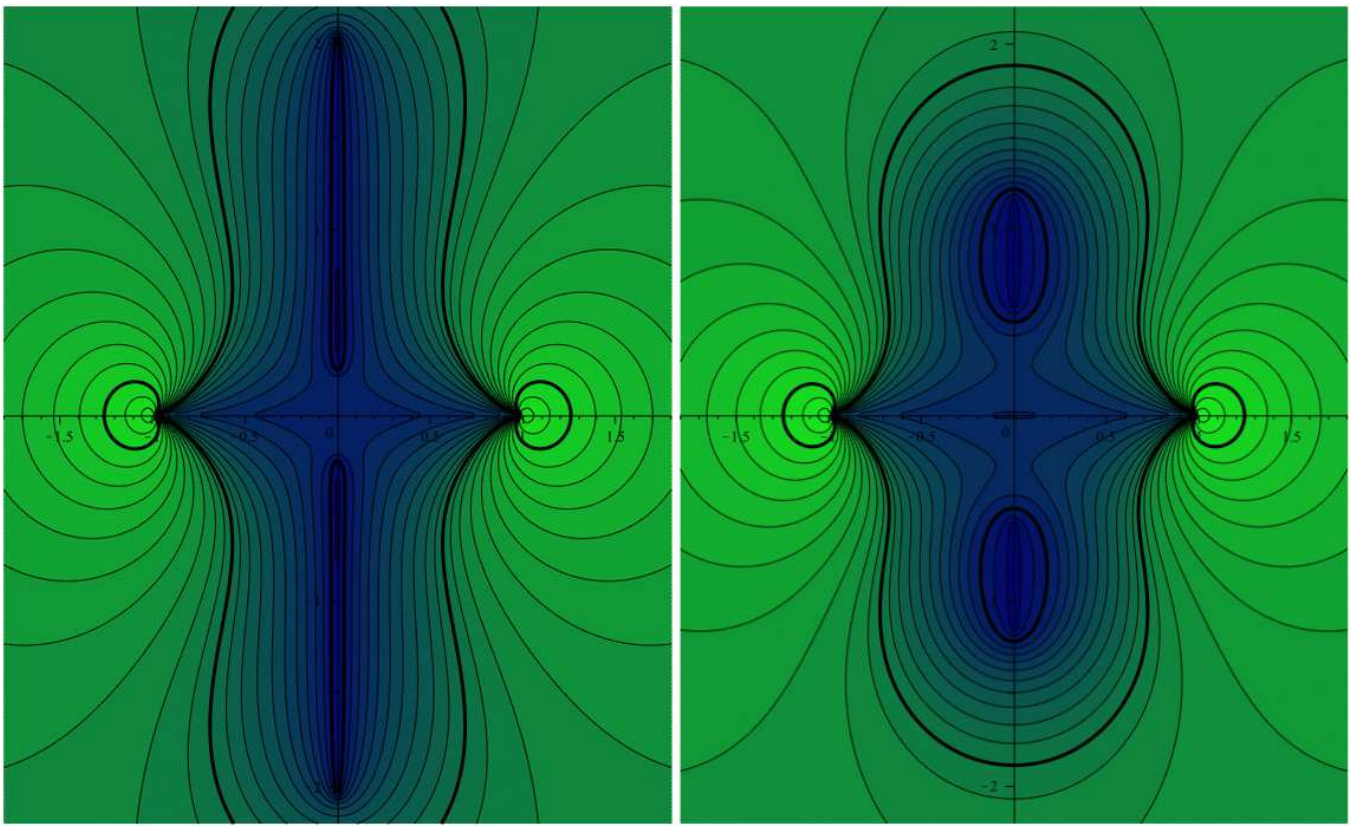}
\caption
{The second sheets (described by $R\!<\!0$) of the same potential superpositions as in figure \ref{potential}, plotted in the same way. The picture differs only in different (positive) sign of the Appell-ring contribution (\ref{nuApp}). In the left plot, ${\cal M}\!=\!3a$, $M\!=\!0.9a$ and $h\!\doteq\!1.1515a$, while in the right plot, ${\cal M}\!=\!a$, $M\!=\!0.3a$ and $h\!\doteq\!0.8713a$ again. The level values range from $-7.5$ to $+7.5$ in the left plot, while within three times weaker values in the right plot. Deep potential is dark blue, while more shallow or even positive one (in regions dominated by the Appell ring) is light green. For comparison with figure \ref{potential}, we have again emphasized the contour $\nu\!=\!-2.86$ in the left plot and the corresponding $\nu\!=\!-0.95$ contour in the right plot (they enclose the black holes quite tightly), but also their counterparts $\nu\!=\!+2.86\,$/$\,\nu\!=\!+0.95$ (which enclose the ring) as well as the $\nu\!=\!0$ contours (the ``largest" of the emphasized ones).}
\label{potential_sheet2}
\end{figure*}

\begin{figure}
\includegraphics[width=\columnwidth]{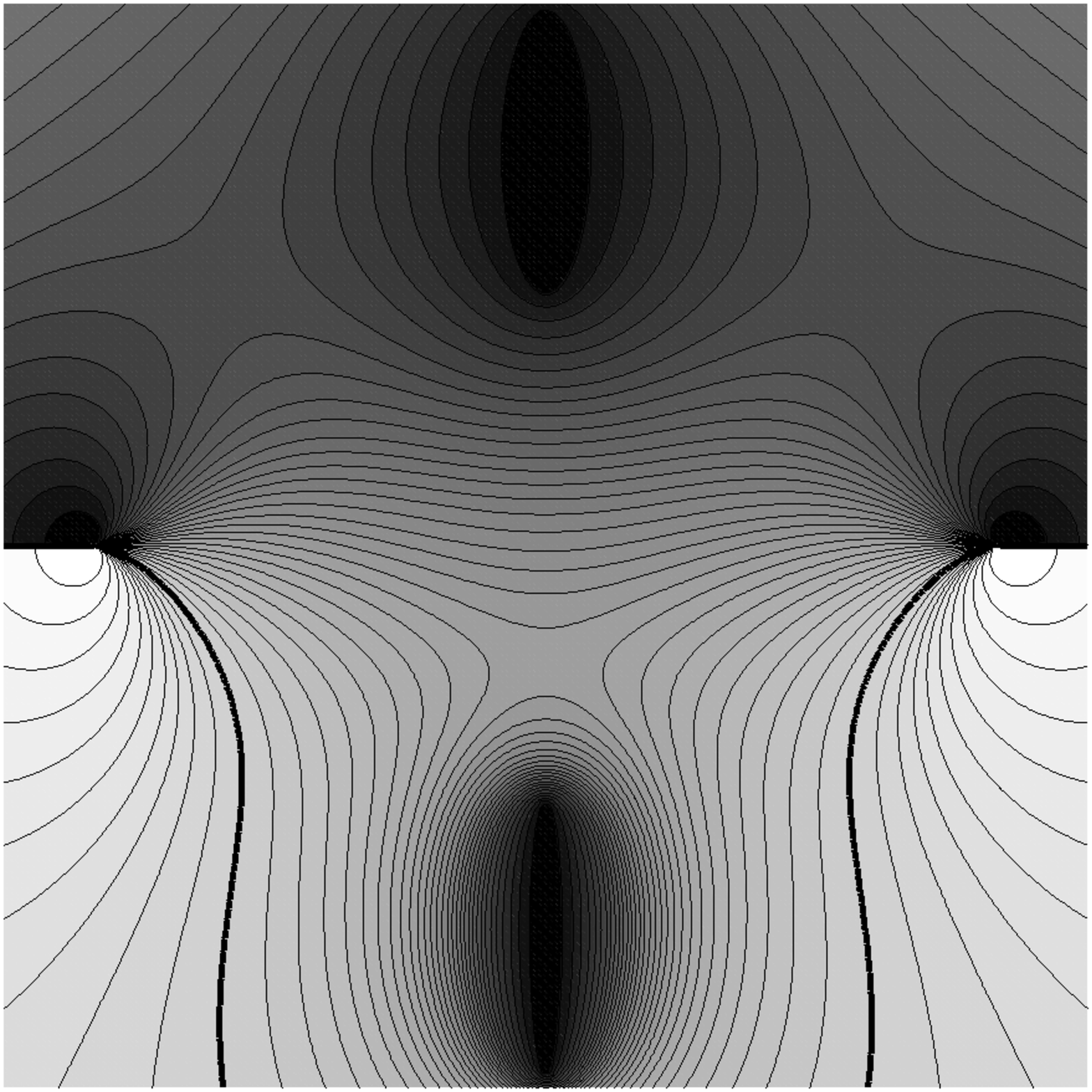}
\caption
{Total potential $\nu$ in the central region of the equilibrium configuration specified by ${\cal M}\!=\!a$, $M\!=\!0.3a$ and $h\!\doteq\!0.8713a$, with the $R\!>\!0$ (top half) and $R\!<\!0$ (bottom half) sheets smoothly connected across the circle $R\!=\!0$ (i.e. $z\!=\!0$, $\rho\!<\!a$). Axes are suppressed, but the radius of the ring is clearly visible (outside it, there is a discontinuity across the equatorial plane), as well as the black holes in both sheets. Low/high potential is black/white, with the $\nu\!=\!0$ contour drawn thick.}
\label{potential-center}
\end{figure}

It is natural to assume that the system is asymptotically flat and that $\lambda\!=\!0$ on the ``outer" parts of the axis $(\rho\!=\!0,|z|\!>\!h+M)$. To check whether the ``inner" part of the axis $(\rho\!=\!0,\,|z|\!<\!h-M)$ can also stay regular under certain conditions, one needs to integrate the equations (\ref{lambda-eqs}) along some path going from the outer axis to the inner one through the vacuum region. Since the black holes are just linear segments $(\rho\!=\!0,\,h-M\leq|z|\leq h+M)$, it is sufficient to take e.g. the path
\begin{widetext}
\[(\rho=0,\,z=h+M+\epsilon) \;\rightarrow\;
  (\rho=\epsilon,\,z=h+M+\epsilon) \;\rightarrow\;
  (\rho=\epsilon,\,z=h-M-\epsilon) \;\rightarrow\;
  (\rho=0,\,z=h-M-\epsilon)\]
in the $\epsilon\!\rightarrow\!0^+$ limit. Integration yields
\[\int\limits_0^\epsilon \lambda_{,\rho}(z=h+M+\epsilon)\,{\rm d}\rho \;
  +\!\!\!\!\int\limits_{h+M+\epsilon}^{h-M-\epsilon}\!\!\!\!\!\lambda_{,z}(\rho=\epsilon)\,{\rm d}z \;
  +\int\limits_\epsilon^0 \lambda_{,\rho}(z=h-M-\epsilon)\,{\rm d}\rho \;
 =\epsilon\,O(\rho\!\rightarrow\!0) \;
  -\!\!\!\!\int\limits_{h-M-\epsilon}^{h+M+\epsilon}\!\!\!\!\!\lambda_{,z}(\rho=\epsilon)\,{\rm d}z \;
  -\epsilon\,O(\rho\!\rightarrow\!0) \,.\]
The integrand of the only contributing part can be found to read
\begin{equation}
  \lambda_{,z}(\rho\!=\!\epsilon,\;h-M\leq z\leq h+M)
  = \frac{2{\cal M}\,(z^2-a^2)}{(z^2+a^2)^2}
    -\frac{2}{z-h+M}
    -\frac{8\,M\,h\,z}{[z^2-(h+M)^2][z^2-(h-M)^2]}
    +O(\epsilon^2)
\end{equation}
along the ``upper" black hole, which yields
\begin{equation}
  \lambda(\rho\!=\!0,\;0\!<\!z\!<\!h-M)
  =-\lim\limits_{\epsilon\rightarrow 0^+}
    \int\limits_{h-M-\epsilon}^{h+M+\epsilon}\lambda_{,z}(\rho\!=\!\epsilon)\,{\rm d}z
  = \frac{4{\cal M}M\,(a^2-h^2+M^2)}{(a^2-h^2+M^2)^2+4a^2 h^2}
   -\ln\frac{h^2}{h^2-M^2} \;.
\end{equation}
\end{widetext}

Hence, the condition for a strut-free equilibrium, $\lambda(\rho\!=\!0,\,0\!<\!z\!<\!h\!-\!M)=0$, reads
\begin{equation}  \label{equilibrium-condition}
  \frac{4{\cal M}M\,(a^2-h^2+M^2)}{(a^2-h^2+M^2)^2+4a^2 h^2}
  =\ln\frac{h^2}{h^2-M^2} \;.
\end{equation}
Firstly, without the ring (${\cal M}\!=\!0$), the only solution is $M\!=\!0$, as expected.
In a non-trivial case, $h\!>\!M\!>\!0$ must hold, hence the right-hand side logarithm is always positive and one obtains an {\em upper} bound for $h$, $h^2\!<\!a^2+M^2$, which obviously says that the black hole must not be too far from the ring in order to be repelled at all. There are two limits within which the equilibrium solution always lies, both being given by ${\cal M}\!\rightarrow\!\infty$ (and $M\!\neq\!0$): the condition can then be satisfied only by $h^2\!\rightarrow\!(M^2+a^2)^-$ or $h\!\rightarrow\!M^+$. The former yields the top boundary (the red dotted hyperbola) and the latter yields the bottom boundary (the red dotted diagonal) in figure \ref{equilibria}.

The figure can be read in various ways of which we suggest the following one: for each ring mass ${\cal M}/a>\!0$, one has a certain equilibrium line in the $(M,h)$ plane which connects two test-particle limits, at $M\!=\!0$, $h\!=\!a$ (where the ring provides an equilibrium location) and at $M\!=\!0$, $h\!=\!0$ (where, in the limit, the energy with respect to infinity of each of the two bodies exactly equals their rest energy, so their mutual attraction there is just as strong as repulsion exerted by the disc $R\!=\!0$). From the behaviour of the test-particle limit, we infer that the repulsion wins to the left of the equilibrium lines (``inside" them), while to the right of the lines (``outside" them) it is the attraction between the two black holes which prevails. Hence, for any positive ${\cal M}/a$ there is a range of $M$ going from zero to a certain positive value, for which {\em two} equilibrium solutions exist; we will choose the larger-$h$ solution which represents the ``expected branch", reducing to the test-particle case $h\!=\!a$ in the $M\!\rightarrow\!0^+$ limit and stable in the $z$ direction (the latter follows from the above fact that inside the equilibrium lines the holes are repelled, whereas outside they are attracted towards each other).

Note a shortcut for derivation of the equilibrium condition, starting from the relation $\lambda_{,z}\!=\!2\nu_{,z}$ which is valid on the horizons and yields $\lambda=2\nu+{\rm const}$ there.\footnote
{Should $\lambda$ be zero at the ``poles" $(\rho\!=\!0,z\!=\!h\pm M)$, it has to read $\lambda=2\nu-2\nu(z\!=\!h\pm M)$, specifically.}
The main benefit of this relation is that it is {\em linear}: since $\nu$ superposes linearly everywhere, this means that specifically on the horizon $\lambda$ does it so, too. Therefore, the $\lambda_{,z}\!=\!2\nu_{,z}$ relation has to also hold separately for each of the three contributions.\footnote
{One might doubt about linearity between $\nu$ and $\lambda$ at the very ``poles" where the functions are singular. However, this singularity is only a coordinate one; in terms of the Schwarzschild-type latitude $\theta$, the horizon-valid relation also reads $\lambda_{,\theta}\!=\!2\nu_{,\theta}$ and is regular everywhere including the poles.}
Consider one of the black holes, say the ``top" one (at $z\!>\!0$). Its own contribution $\nu_{\rm Schw}(z\!-\!h)$ naturally satisfies the relation, however diverging on the horizon, so one is left with the potentials from the Appell ring and from the second black hole which are finite there. The function $\lambda$ has the same value on both parts of the axis (thus on all three actually) if it satisfies this at the horizon ``poles" $(\rho\!=\!0,z\!=\!h\pm M)$, which is now clear to hold if the sum of the ``external" potentials,
$\left[\nu_{\rm App}(z)+\nu_{\rm Schw}(z\!+\!h)\right]$,
assumes the same value at both poles (this is a well known condition, see \cite{GerochH-82}). Substituting
\begin{align*}
  \nu_{\rm App}(\rho\!=\!0,z) &=-\frac{{\cal M}\,z}{z^2+a^2} \;, \\
  \nu_{\rm Schw}(\rho\!=\!0,z+h) &=\frac{1}{2}\,\ln\frac{z+h-M}{z+h+M}
\end{align*}
into the above, one really arrives at the condition (\ref{equilibrium-condition}).

For the equilibrium system, the first term in (\ref{explambda,circle}) is $e^\lambda(\rho\!=\!0,z\!=\!0)\!=\!1$, so the integrand of the integral $\int_0^a e^{\lambda-\nu}(z\!=\!0)\,{\rm d}\rho\,$ starts from
\[e^{\lambda-\nu}(\rho\!=\!0,z\!=\!0)=e^{-\nu}(\rho\!=\!0,z\!=\!0)=\frac{h+M}{h-M}\geq 1\]
and falls to zero at the ring. The behaviours of the ring's proper circumference and proper radius are best illustrated if plotted against each other, see figure \ref{radius-circumference}. The curves are parametrized by the black-hole mass $M$ (the ranges correspond only to vertically stable, ``upper" branches of the ${\cal M}\!>\!0$ equilibria from figure \ref{equilibria} again), with Euclidean relation added (dotted blue line) for reference. The circumference--radius dependence would indicate negative curvature for small $M$, while positive curvature for larger $M$.

As a basic illustration of what field the system generates, figure \ref{potential} shows the meridional equipotentials for two examples of equilibrium configurations, one rather ``heavy" (with ${\cal M}\!=\!3a$, $M\!=\!0.9a$ and $h\!\doteq \!1.1515a$) and the other three-times lighter (with ${\cal M}\!=\!a$, $M\!=\!0.3a$ and  $h\!\doteq\!0.8713a$). Quite a steep gradient of the potential is visible between the ring's centre and the black-hole horizons, mainly in the heavier case, which promises to yield interesting deformation of the field and curvature {\em inside} the black holes.
We are also appending the plots of the second (negative-$R$) sheets for the same two superpositions, see figure \ref{potential_sheet2}. The potentials appear there the same, just the Appell-ring contribution (\ref{nuApp}) has an opposite (positive) sign. Hence, in this second sheet, the Appell ring is repulsive (effectively, its mass is negative), only the central circle $R\!=\!0$ (actually the whole region $0\!\geq\!R\!>\!-a|\!\cos\vartheta|$) is ``attractive", and ``mirror images" of the black holes are present at the same locations as in the $R\!>\!0$ sheet. Figure \ref{potential-center} confirms that the potential $\nu$ continues smoothly across the central $R\!=\!0$ circle. The region $R\!<\!0$ is of course irrelevant (non-existing) if one accepts the interpretation with negative-density mass layer on $R\!=\!0$.

It is worth remarking that when adopting the smooth-central-circle view (the double-sheeted topology), any of the black holes feels not only its symmetrical counter-part lying in the same sheet yet ``behind" the $R\!=\!0$ throat (leading to the other sheet), but also -- through the $R\!=\!0$ throat -- its counter-part lying in the {\em other} sheet. Actually, there clearly exist geodesics following the axis of symmetry and connecting the two black holes in the opposite sheets. However, the field at a given location and the equilibrium configurations are independent of the adopted interpretation, so it would have little sense, in case of the second interpretation, to try to split the attractive part of the effect on a given black hole into the above contributions.

\subsection{Negative ring mass?}

When speaking of black holes, it is automatically assumed that their mass is positive. When speaking of sources like the Appell ring (but also e.g. the Kerr source), however, the choice is not that clear, because what is called ``positive" mass there means that, for test particles at rest, the ring is attractive at $R\!>\!a|\!\cos\vartheta|$ (and also at $0\!>\!R\!>\!-a|\!\cos\vartheta|$), whereas it is repulsive in the remaining regions ($R\!<\!-a|\!\cos\vartheta|$ and $0\!<\!R\!<\!a|\!\cos\vartheta|$). Like in the case of Kerr solution, the ``positive"-mass choice is standardly connected with choosing $R\!>\!0$ (i.e. the region where, at medium and large radii, the field is attractive then) as the relevant half of space-time. With the opposite choice, it would not be unreasonable to also consider negative masses.

For ${\cal M}\!<\!0$ (and $M\!>\!0$), the equilibrium condition (\ref{equilibrium-condition}) can only be satisfied if $h^2\!>\!M^2+a^2$. Region of equilibria is again bound by the limit ${\cal M}\!\rightarrow\!-\infty$ which can only hold for $h^2\!\rightarrow\!(M^2+a^2)^+$. In figure \ref{equilibria}, we also include parts of several curves of equilibria corresponding to ${\cal M}\!<\!0$, but in the other plots as well as in the rest of the paper, we restrict ourselves to the ${\cal M}\!>\!0$ case. Note in passing that the equilibria obtained with ${\cal M}\!<\!0$ are generally given by larger $h/M$ than those obtained with ${\cal M}\!>\!0$, so one may expect that the effect of the ring on the black hole is {\em weaker} (and thus less interesting) in those cases.

\section{Simple properties of the horizons}
\label{horizons}

Let us now check basic dimensions of the black holes.
The proper distance between the ring's centre at $(\rho\!=\!0,z\!=\!0)$ and the horizons of the black holes, measured along the $z$ axis, amounts to
\begin{align}
  &\int\limits_0^{h-M}\sqrt{g_{zz}(\rho\!=\!0)}\;{\rm d}z
   =\int\limits_0^{h-M}e^{-\nu(\rho=0)}\,{\rm d}z= {} \nonumber \\
  &=\int\limits_0^{h-M}\sqrt{\frac{z^2-(h+M)^2}{z^2-(h-M)^2}}\;
                       \exp\frac{{\cal M}z}{z^2+a^2}\;{\rm d}z \,,
\end{align}
where we have supposed to deal with the equilibrium configuration, corresponding to $\lambda(\rho\!=\!0)\!=\!0$ along the integration path. The results have been given in figure \ref{equilibria}.

\begin{figure*}
\includegraphics[width=0.85\textwidth]{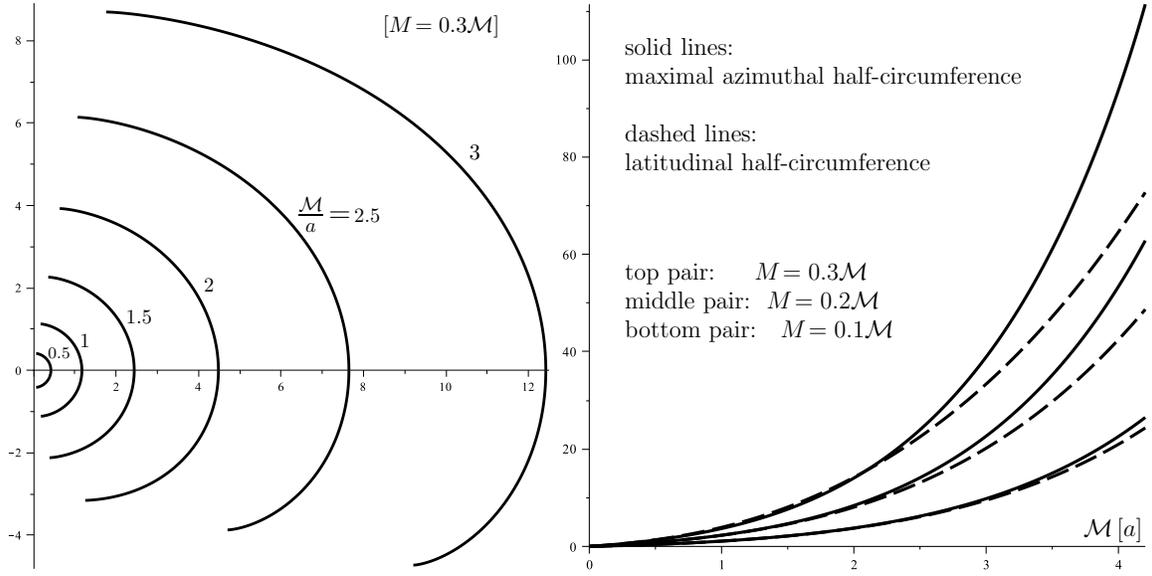}
\caption
{{\it Left:} Isometric picture of the horizon's meridional section within the Euclidean plane: azimuthal circumferential radius is plotted along the horizontal axis, while proper distance goes along the outlines. Several equilibrium configurations are included (taken from upper branches of figure \ref{equilibria}), namely those with ${\cal M}/a=0.5$, 1, 1.5, 2, 2.5 and 3, and with $M\!=\!0.3{\cal M}$, as indicated; the corresponding equilibrium $z$-positions are $h/a=0.8408$, 0.8713, 0.9202, 0.9851, 1.0632 and 1.1515, respectively. The outlines cannot be drawn in whole when the Gauss curvature of the horizon is negative in a certain region including the pole; this typically happens on the side towards the central ring. The vertical axis is adjusted so that its zero correspond to the ``parallel of latitude" with maximal azimuthal circumference. 
{\it Right:} Comparison of the black-hole azimuthal and latitudinal circumferences in dependence on the Appell-ring mass ${\cal M}$, computed for the equilibrium configurations again. Solid lines show half of the maximal azimuthal ($\phi$) circumference while dashed lines show half of the latitudinal circumference (computed as the length of the horizons along the $z$-axis in Weyl coordinates). The respective pairs of curves correspond (from bottom to top) to the black-hole masses $M\!=\!0.1{\cal M}$, $M\!=\!0.2{\cal M}$ and $M\!=\!0.3{\cal M}$.}
\label{BH-shape}
\end{figure*}

The proper azimuthal circumference of the horizons is found to read
\begin{align}
  & 2\pi\lim\limits_{\rho\rightarrow 0}\sqrt{g_{\phi\phi}(h\!-\!M<|z|<h\!+\!M)}= {} \nonumber \\
  & = 2\pi\lim\limits_{\rho\rightarrow 0}(\rho\,e^{-\nu})|_{h-M<|z|<h+M}= {} \nonumber \\
  & = 4\pi\;\frac{\sqrt{(|z|+h+M)(|z|-h+M)(h+M-|z|)}}{\sqrt{|z|+h-M}}\,\times {} \nonumber \\
  & \quad \times \exp{\frac{{\cal M}\,|z|}{a^2+z^2}} \,.
\end{align}
Although the horizons are not represented reasonably in the Weyl coordinates, it is clear that their azimuthal circumference depends on $z$ and that in our composed system it is {\em not} likely to be maximal exactly at $z\!=\!h$. This will become clear in figure \ref{BH-shape}.

The ``proper length" of the horizons along the $z$-axis -- actually $1/2$ of their proper meridional circumference if imagining in more appropriate coordinates -- is given by
\[\int_{h-M}^{h+M}\sqrt{g_{zz}(\rho\!=\!0)}\,{\rm d}z=\int_{h-M}^{h+M}e^{\lambda-\nu}(\rho\!=\!0)\,{\rm d}z \,,\]
where $\lambda\!=\!2\nu-2\nu(z\!=\!h+M)$. Consider again the ``top" black hole. Although $\nu_{\rm Schw}(z\!-\!h)$ is of course singular there, from linear addition of $\nu$ (and thus of $\lambda$) we find easily that at its horizon (BH+)
\begin{align}
  \nu_{\rm BH+}     =& \; \ln\frac{\rho}{2\sqrt{(z-h+M)(h+M-z)}}+ {} \nonumber \\
                     & \; +\ln\sqrt{\frac{z+h-M}{z+h+M}}
                          -\frac{{\cal M}\,z}{a^2+z^2}
                          +O(\rho^2) \,, \\
  \lambda_{\rm BH+} =& \; \ln\frac{M\,\rho}{(z-h+M)(h+M-z)}
                          +O(\rho^2) \,,
\end{align}
hence the logarithmic divergence (brought by the given black hole itself) cancels out,
\begin{align}
  &\exp(2\lambda_{\rm BH+}\!-2\nu_{\rm BH+})= \nonumber \\
  &=\frac{4M^2\,(z+h+M)}{(z\!-\!h\!+\!M)(h\!+\!M\!-\!z)(z\!+\!h\!-\!M)}\;
    \exp\frac{2{\cal M}\,z}{a^2\!+\!z^2} \;.
\end{align}

\begin{figure*}
\includegraphics[width=\textwidth]{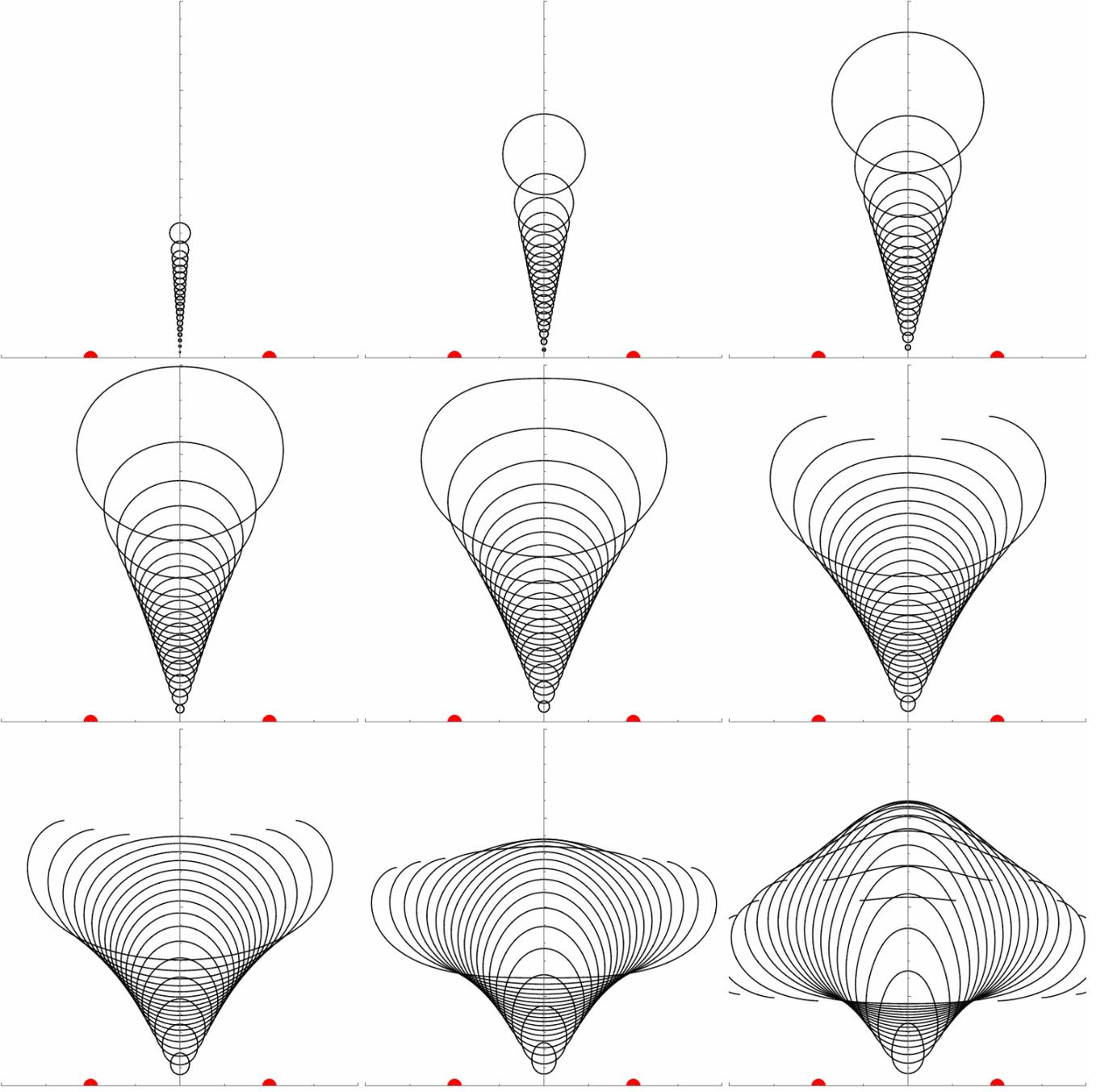}
\caption
{Isometric representation of the top-black-hole horizon for a number of different parameter combinations. From top to bottom and from left to right, the $M/h$ is set to 0.05, 0.15, 0.25, \dots, 0.85, and each of these nine plots contain 19 different horizon outlines obtained for $h/a=$ 0.05, 0.10, 0.15, \dots, 0.95 (and the ring mass fixed by the condition for strut-free equlibrium). The ring radius $a$ is used as a length unit, so the ring section is on the horizontal axis at $\pm 1$ (red dots). The plot dimensions are $\langle -2,+2\rangle \times \langle 0,4\rangle$. For higher values of $M/h$, some of the horizons are no longer completely embeddable, with the problem first occurring at the symmetry axis (vertical axis of the plots).}
\label{shape-05-85_1}
\end{figure*}

\begin{figure*}
\includegraphics[width=\textwidth]{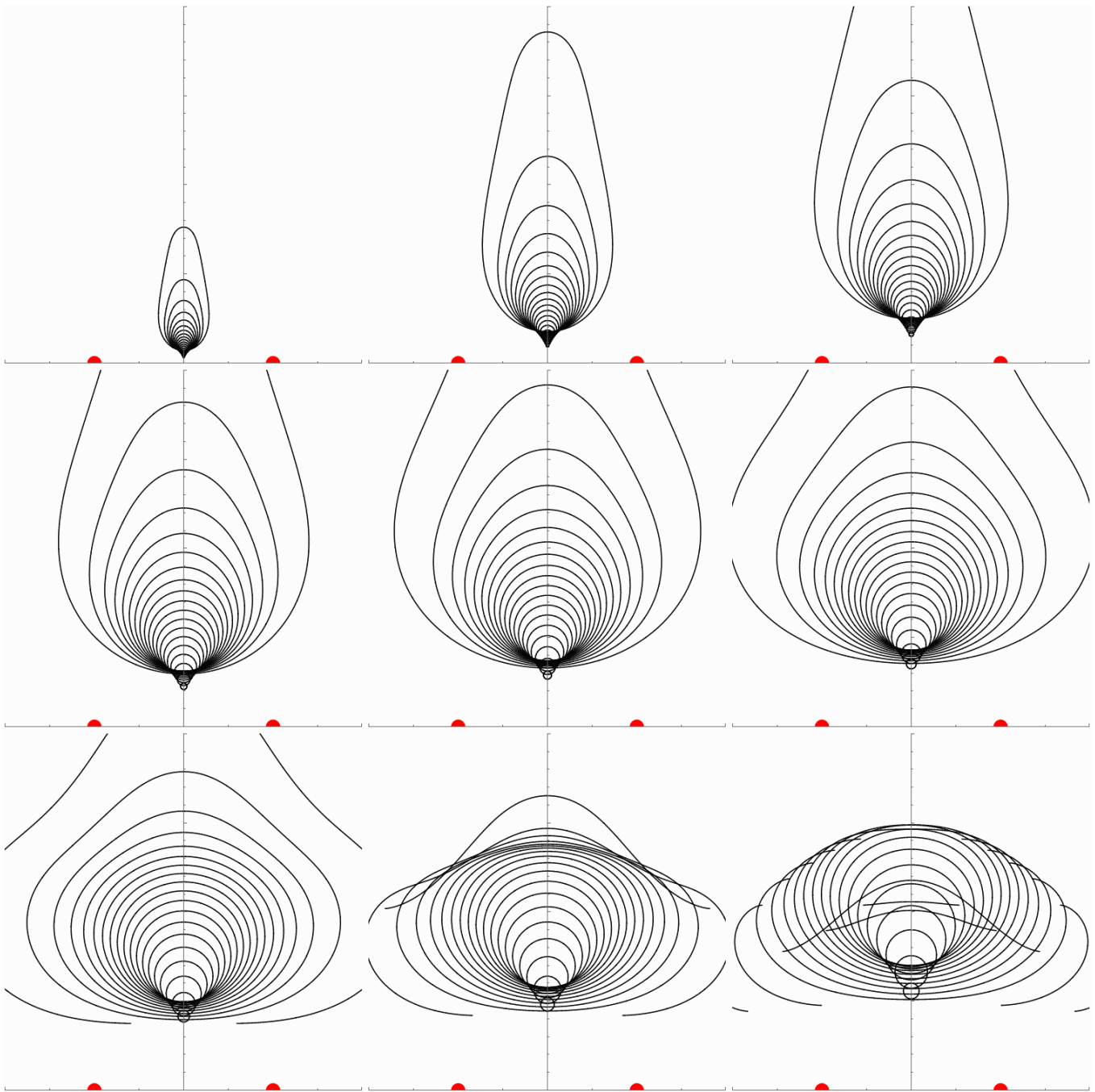}
\caption
{The same sequence of plots as in figure \ref{shape-05-85_1}, but now computed for the counterpart of the black hole lying on the second sheet of space-time (at $R\!<\!0$).}
\label{shape-05-85_2}
\end{figure*}

\begin{figure}
\includegraphics[width=\columnwidth]{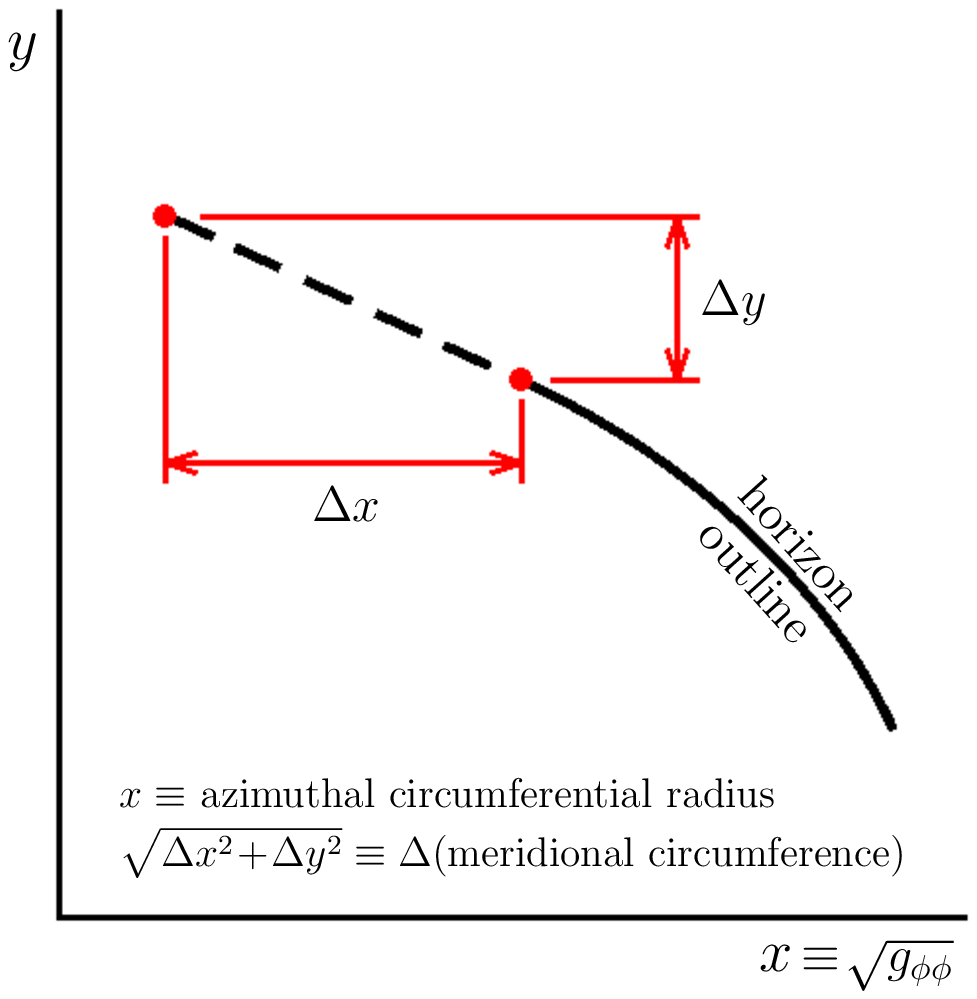}
\caption
{A scheme of the meridional outline of the horizon, as showing properly its azimuthal circumferential radius ($x$ coordinate) as well as its meridional circumference (along the outline). For a given step $\Delta x$, the corresponding step in the meridional circumference has to at least equal $\Delta x$, otherwise the curve cannot be drawn in $\mathbb{E}^2$. The spherical radius of a given horizon point may loosely be understood as its latitudinal circumferential radius.}
\label{shape-scheme}
\end{figure}

Knowing the proper ``vertical" (in fact latitudinal) length along the horizon and its azimuthal circumferential radius in dependence on $z$, we can plot its ``true shape", namely its meridional-section outline plotted isometrically within the Euclidean plane; this is done in figure \ref{BH-shape} (left-hand plot) for several values of the masses ${\cal M}$ and $M$. The figure shows that with increasing masses the horizon gets oblate, with flattening mainly occurring on the side towards the central ring. The right-hand plot of figure \ref{BH-shape} confirms that with increasing masses the maximal azimuthal circumference grows more quickly than the latitudinal one, consistently with the flattening. The horizon however cannot be drawn in whole, which is a familiar problem in case when the Gauss curvature turns negative in certain regions including the axis, mainly on the side of the repulsive ring. These regions grow when masses of the sources are enlarged.

To find the Gauss curvature of the horizon means to compute (half of) the Ricci scalar for the metric of the 2D horizon $\{t\!=\!{\rm const},\,\rho\!=\!0,\,h-M\leq z\leq h+M,\,\phi\}$ which in the Weyl coordinates appears as
\begin{widetext}
\begin{align}
  {\rm d}\sigma^2
  &\eqH \lim\limits_{\rho\rightarrow 0^+}
        \left(\rho^2 e^{-2\nu}{\rm d}\phi^2 + e^{2\lambda-2\nu}{\rm d}z^2\right)_{h-M<z<h+M}
        \nonumber \\
  &= 4\,\frac{z+h+M}{z+h-M}\,\exp\frac{2{\cal M}z}{a^2+z^2}
     \left[(z-h+M)(h+M-z)\,{\rm d}\phi^2+\frac{M^2\;{\rm d}z^2}{(z-h+M)(h+M-z)}\right].
\end{align}
\end{widetext}
We can remind the connection between the Gauss curvature of some surface and the possibility of its isometric embedding. Actually, the difficulty with embedding arises where the azimuthal circumferential radius (the ``$x$"-coordinate of the surface) changes faster than its latitudinal circumference (whose infinitesimal step should be $\sqrt{\Delta x^2+\Delta y^2}\,$) -- cf. Fig. \ref{shape-scheme}. The change of the azimuthal circumferential radius being given by
\[\sqrt{g_{\phi\phi}(z+{\rm d}z)}-\sqrt{g_{\phi\phi}(z)}\simeq (\sqrt{g_{\phi\phi}})_{,z}{\rm d}z\]
and the latitudinal-circumference step by $\sqrt{g_{zz}}\,{\rm d}z$, the embedding condition reads
\[\left|(\sqrt{g_{\phi\phi}})_{,z}\right|\leq\sqrt{g_{zz}} \;.\]
But the derivative of the ratio of these two is proportional to the Gauss curvature of the surface (see \cite{BasovnikS-16}, equation (62)),
\begin{align}
  &\frac{\partial}{\partial z}\left[\frac{(\sqrt{g_{\phi\phi}})_{,z}}{\sqrt{g_{zz}}}\right]
   =-\sqrt{g_{\phi\phi}g_{zz}}\;\frac{^{(2)}\!R}{2} \,,
\end{align}
where $^{(2)}\!R$ is the Ricci scalar of the given 2D surface.
Hence, if
\[{\rm emb}:=\frac{(\sqrt{g_{\phi\phi}})_{,z}}{\sqrt{g_{zz}}}\]
is everywhere a decreasing function of $z$, there should be no problem with embedding. The opposite statement is more subtle. In our case -- the spheroidal and axisymmetric surface -- we know, however, that for the poles to be ``elementary flat", the circumferential radius has to change there exactly as the proper latitudinal (or $z$-) circumference, namely that
\[{\rm emb}(z=h+M)=1 \,, \quad {\rm emb}(z=h-M)=-1 \,.\]
Therefore, the emb-function has to be ``more decreasing than increasing" when going from the top pole to the bottom one. It is still possible that it increases within some segment -- and such a circumstance need not necessarily imply that the embedding is impossible, not even that it is impossible {\em there}. The problem only arises if the negative-$^{(2)}\!R$ region involves the axis, i.e. -- in the axisymmetric case -- if the region is simply connected. Even in such a case, however, it is not so that the embedding would fail exactly in the region where $^{(2)}\!R\!<\!0$. (This is standardly being illustrated on a saddle surface whose Gauss curvature is everywhere negative, yet still a certain part of it {\em can} be drawn.)

In order to even more illustrate the change of the horizon geometry with parameters, we add figures \ref{shape-05-85_1} and \ref{shape-05-85_2} which show, for any of the black holes placed in the first space-time sheet (at $R\!>\!0$) and for its counter-part lying in the second sheet (at $R\!<\!0$), an isometric embedding of the horizon in the Euclidean space. Each of the figures actually contain 9 plots which correspond to 9 different values of the ratio $M/h$ (from 0.05 to 0.85, by 0.10), and each of the plots contain 19 horizon profiles obtained for 19 different ratios $h/a$ (from 0.05 to 0.95, by 0.05). For each specific choice of the parameters $a$ (which serves as the length unit), $M$ and $h$, the ring mass ${\cal M}$ is determined by the equilibrium condition (\ref{equilibrium-condition}), so it is different for each of the horizons. The embeddings are made so that the plots show correctly the proper distance between the ring center and the ``bottom'' point of the horizons, and of course the horizon shape. For higher values of the parameters (which generally correspond to larger strain the black hole is subjected to), some of the horizons can no longer be globally embedded, so only parts of them are drawn.

\section{Geometry in the black-hole exterior}
\label{BH-exterior}

\begin{figure}
\includegraphics[width=0.99\columnwidth]{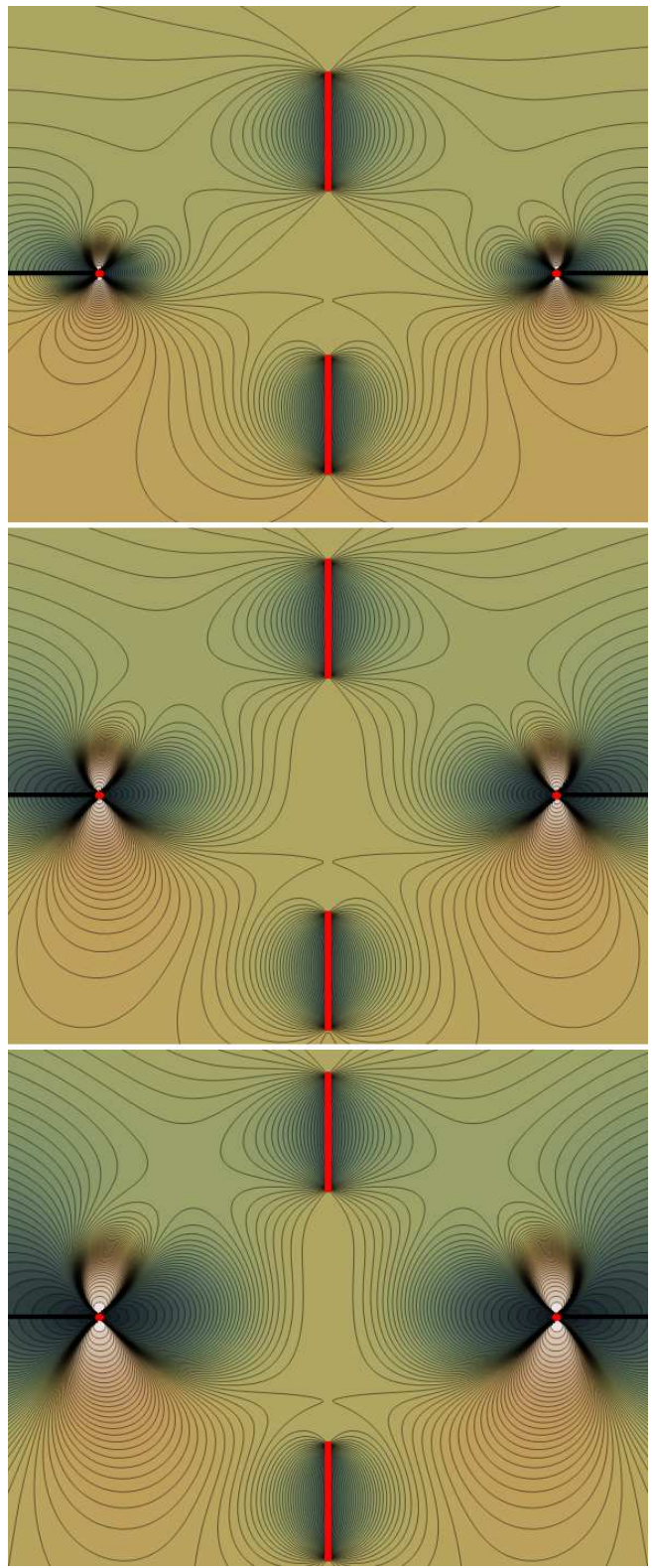}
\caption
{Meridional plot of total $\lambda$ for three strut-free configurations: $({\cal M},h,M)$ $=$ $(0.7,0.753,0.3)$ \!/\! $(1.5,0.936,0.3)$ \!/\! $(2.5,0.982,0.3)$ (top/middle/bottom) in the units of $a$. Red segments are horizons (fixing the symmetry axis), red dots are ring sections (fixing the equatorial plane). The $R\!>\!0$/$R\!<\!0$ sheets are shown above/below the ring plane, like in figure \ref{potential-center}.}
\label{lambda-out}
\end{figure}

\begin{figure*}
\includegraphics[width=\textwidth]{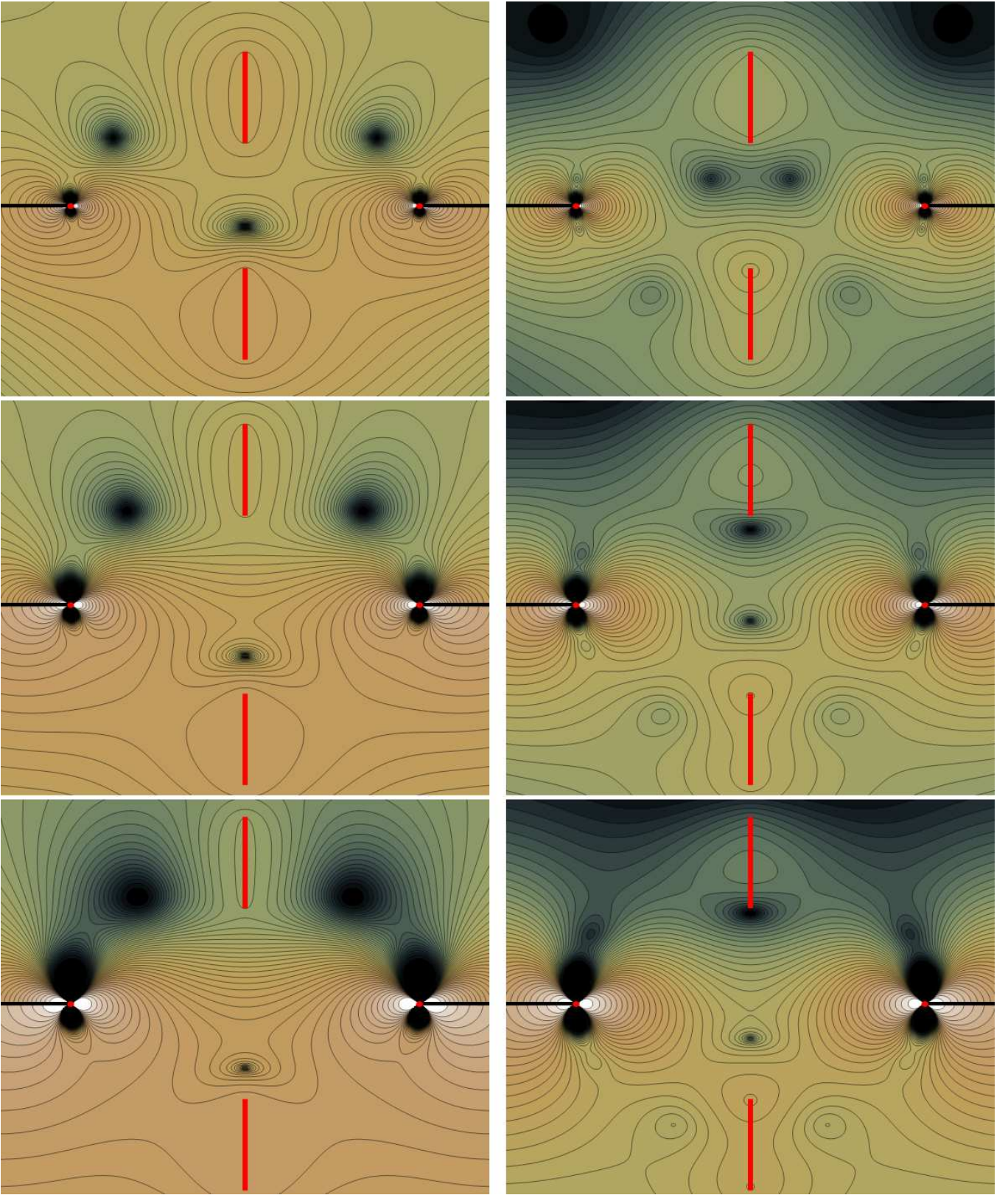}
\caption
{Meridional plot of the acceleration and Kretschmann scalars $\kappa^2$ and $K$ (left/right column) for the same three configurations as in figure \ref{lambda-out}. All quantities are positive in both figures, with brown/dark green indicating large/small values.}
\label{acc-Kretschmann-out}
\end{figure*}

Outside of the black holes, the properties of the system can be analyzed in the Weyl coordinates. Since the potential (thus lapse-function) contours have already been exemplified, for the strut-free configuration, in figures \ref{potential}--\ref{potential-center}, we only add illustrations of the second metric function $\lambda$ (figure \ref{lambda-out}), and of acceleration and curvature invariants (figure \ref{acc-Kretschmann-out}). The meridional contours of the quantities are drawn for three examples of the strut-free equilibrium. The values are everywhere positive, with brown colour indicating heights and dark green (asymptotically black) indicating valleys. The geometry is seen to be strongly deformed in the vicinity of the sources, mainly close to the ring.

\section{Geometry in the black-hole interior}
\label{BH-interior}

In order to describe the black-hole interior, we transform from the Weyl coordinates $\rho$, $z$ to the Schwarzschild-type coordinates $r$, $\theta$. Adapting the latter to the top/bottom black hole (contributing by the potential $\nu_{\rm Schw}(z\!\mp\!h)$), thus taking $\rho\!=\!0$ and $z\!=\!\pm h$ as origin, such a transformation reads
\begin{equation}
  \rho=\sqrt{r(r-2M)}\,\sin\theta, \quad
  z\mp h=(r-M)\,\cos\theta.
\end{equation}

\subsection{Free fall to the singularity}

The first obvious question which arises is whether the central singularity of the black holes is shifted off its symmetric position (central to the horizon). It is clear from symmetry that the singularity remains on the axis ($\sin\theta\!=\!0$), while its ``actual" vertical position can be deduced from the time of a free fall from the horizon. Since the limit case of a test particle (with rest mass $m$) dropped from rest from the horizon corresponds to a vanishing conserved energy with respect to infinity,
\[0=E:=-p_t=-g_{tt}mu^t \,,\]
one sees that $u^t$ has to vanish as well along the whole geodesic starting on the horizon and ending at the singularity. Restricting to the geodesics following the symmetry axis ($\sin\theta\!=\!0$), for which both $u^\theta\!=\!0$ and $u^\phi\!=\!0$, the four-velocity normalization thus reduces to
\[g_{rr}({\rm axis})(u^r)^2=-1
  \quad \Rightarrow \quad
  \frac{{\rm d}r}{{\rm d}\tau}=-\frac{1}{\sqrt{-g_{rr}({\rm axis})}}\]
and the total time of flight is found by integrating this from $r\!=\!2M$ to $r\!=\!0$.
Substituting $g_{\rho\rho}\!=\!g_{zz}\!=\!e^{2\lambda-2\nu}$ for the Weyl-metric components, one finds that
\begin{align*}
  g_{rr}&=\left(\frac{\partial\rho}{\partial r}\right)^{\!2} g_{\rho\rho}+
          \left(\frac{\partial z}{\partial r}\right)^{\!2} g_{zz} \\
        &=\frac{(r-M)^2-M^2\cos^2\theta}{r(r-2M)}\,e^{2\lambda-2\nu} \,.
\end{align*}
On the axis ($\sin\theta\!=\!0$), one has $\lambda\!=\!0$ in order to ensure local flatness of the $\{t\!=\!{\rm const},\,r\!=\!{\rm const}\}$ surfaces there (this has to hold below the horizon as well), and the pre-factor $\frac{(r-M)^2-M^2\cos^2\theta}{r(r-2M)}$ reduces to unity, so it is sufficient to compute
\[\nu = \nu_{\rm App}(z)+\nu_{\rm Schw}(z\!-\!h)+\nu_{\rm Schw}(z\!+\!h)\]
for ($\sin\theta\!=\!0\,\Rightarrow$) $\rho\!=\!0$ and $z\!=\!h\pm(r-M)$, where the plus/minus sign comes from $\cos\theta\!=\!\pm 1$ and applies respectively for the fall from the top/bottom pole of the black hole. At $\rho\!=\!0$, one has
\begin{align}
  &\nu_{\rm App}(z) =-\frac{{\cal M}z}{z^2+a^2}
                    =-\frac{{\cal M}\,[h\pm(r-M)]}{[h\pm(r-M)]^2+a^2} \;, \\
  &\nu_{\rm Schw}(z\!-\!h) =\frac{1}{2}\,\ln\left(1-\frac{2M}{r}\right), \\
  &\nu_{\rm Schw}(z\!+\!h) =\frac{1}{2}\,\ln\frac{z+h-M}{z+h+M}= \nonumber \\
            & \hspace*{18mm} =\frac{1}{2}\,\ln\frac{2h\pm(r-M)-M}{2h\pm(r-M)+M} \,,
\end{align}
hence
\begin{align}
  &-g_{rr}(\cos\theta\!=\!\pm 1)=-e^{-2\nu(\cos\theta=\pm 1)}= \nonumber \\
  &= \frac{r}{2M\!-\!r}\,\frac{2h\pm(r\!-\!M)+M}{2h\pm(r\!-\!M)-M}\,
     \exp\frac{2{\cal M}\,[h\pm(r\!-\!M)]}{[h\pm(r\!-\!M)]^2+a^2} \,.  \label{-grr}
\end{align}

\begin{figure}
\includegraphics[width=\columnwidth]{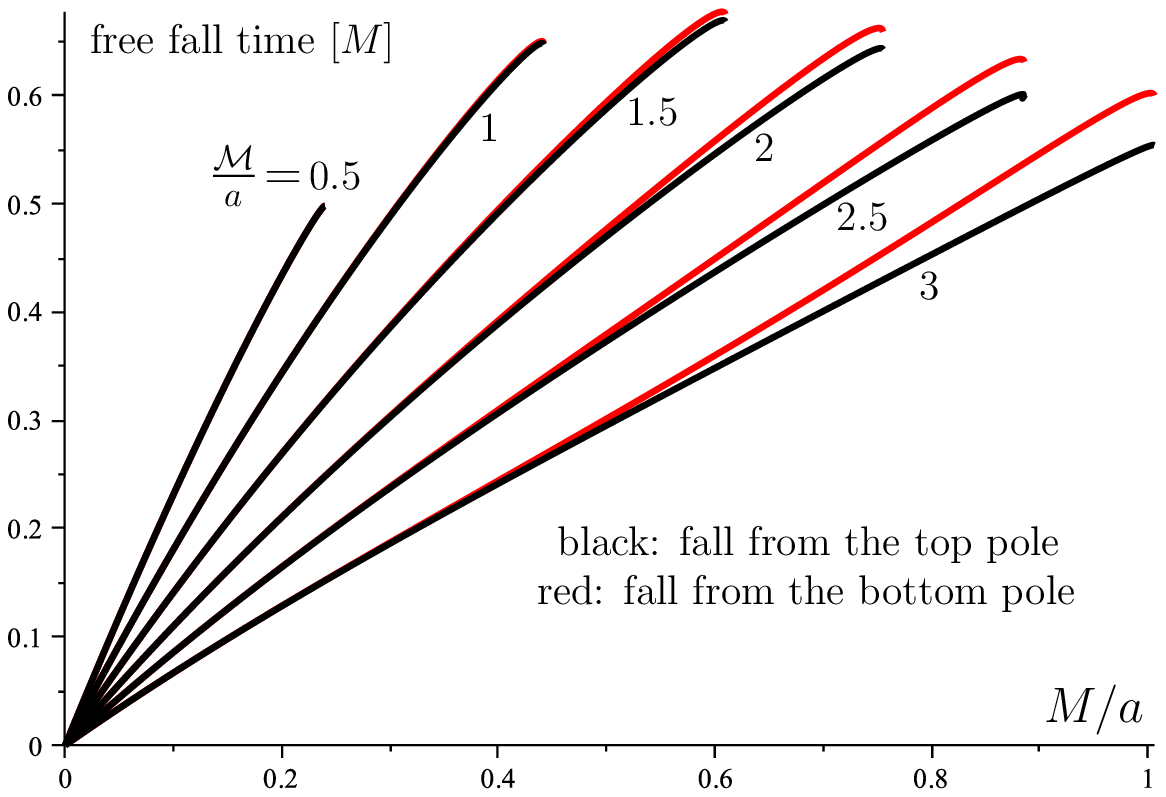}
\caption
{Time of free fall from rest, along the symmetry axis, from the horizon poles to the black-hole singularity. Fall times from the pole which is remote/close to the central Appell ring are shown by black/red lines, for several values of the ring mass ${\cal M}/a$ (as indicated at the lines) and in dependence on the black-hole mass $M/a$. For given ${\cal M}/a$ and $M/a$, the black-hole position $h$ is chosen so as to correspond to the ``vertically" stable branch of the equilibrium configurations described in section \ref{equilibrium-configurations} and illustrated in figure \ref{equilibria}. It is seen that the time of free fall from the side of the Appell ring is slightly longer, but the difference is not large, mainly when the participating masses (${\cal M}$, $M$) are small.}
\label{freefall-time}
\end{figure}

The integration of $(-g_{rr})^{-1/2}$ given by this expression clearly yields different results for the top/bottom signs, which means that the times of free fall from the top/bottom pole of the (top) black hole are different. In a single-Schwarzschild limit, one is left just with the first fraction and obtains the well known value
\[\Delta\tau_{\rm Schw}=\int_0^{2M}\sqrt{\frac{2M}{r}-1}\;{\rm d}r=\pi M.\]
Needless to say, the second fraction of (\ref{-grr}) is due to the second black hole and the exponential is due to the Appell ring. Figure \ref{freefall-time} shows how the result depends on the mass of the ring ${\cal M}$ and on that of the black holes ($M$); it mainly shows that there is only a small difference between the free-fall time from the top and from the bottom pole of the horizon, in particular, that from the pole closer to the Appell ring is slightly longer. This indicates that the external source does not affect the black-hole interior very strongly on the level of field intensity.

\subsection{Describing the black-hole interior}

In \cite{BasovnikS-16}, dealing with a black hole surrounded by a concentric Bach-Weyl ring, we managed to extend the analysis into the black-hole interior (which is not covered by the original Weyl coordinates) along a limit case of null geodesics starting tangentially from the horizon and reaching the singularity after making an arc of exactly one $\pi$ in angular (latitudinal) direction. Here we will proceed in a similar manner.

More precisely, there are two straightforward ways how to extend below horizon.
First, for a function which is analytic on the horizon ($\rho\!=\!0$, $|z|\!\leq\!M$) and which can be extended to a holomorphic function on the whole plane ($0\!\leq\!\rho\!\leq\!{\rm i}z$, $|z|\!\leq\!M$), with $\rho$ being pure imaginary now, one can use the standard transformation (\ref{Schw-Weyl}), with $r$ and $\theta$ representing Schwarzschild-type coordinates {\em adapted to the given black hole}, i.e. having their origin at the respective singularity, and now assuming values below the given horizon ($r\!<\!2M$). This approach is suitable for the external potential $\nu_{\rm ext}$ (represented by the Appell ring and by ``the other'' black hole in the present paper) which, in contrast to the potential of the given black hole itself, is typically analytic at its horizon. One can in particular compute the external potential by integral
\begin{equation}
  \nu_{\rm ext}(\rho,z)
  =\frac{1}{\pi}\int_0^\pi \nu_{\rm ext}(0,z\!-\!{\rm i}\rho\cos\alpha)\,{\rm d}\alpha
\end{equation}
which yields a holomorphic result if $\nu_{\rm ext}$ is real analytic on the $\rho\!=\!0$ axis. Then, the complete interior solution for $\nu$ is obtained, like elsewhere, by adding the $\nu_{\rm ext}$ result to the known $\nu_{\rm Schw}$ generated by the given black hole. Specifically in the configuration we consider here, the potential of the Appell ring as well as that due to the black-hole counterpart lying on the other side of the ring are regular and analytic everywhere inside the given black hole, so they can be extended there holomorphically without problem.

\begin{figure}
\includegraphics[width=0.8\columnwidth]{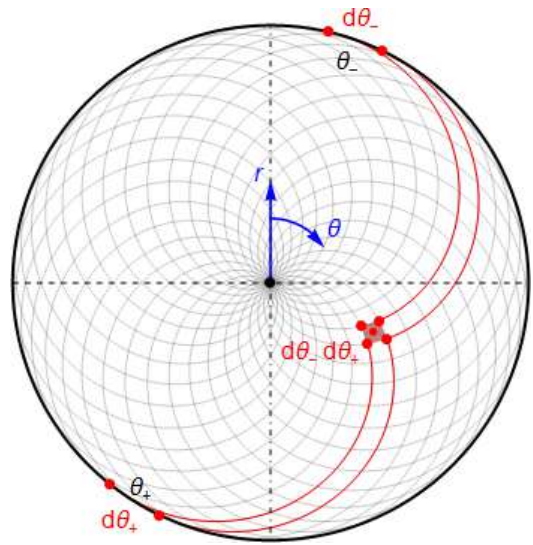}
\caption
{We denote by $\theta_+$, $\theta_-$ the latitudes on the horizon from where the null geodesics start tangentially and then ``counter-inspiral'' towards the singularity, crossing each other at some $(r,\theta)$. The angles are used as coordinates covering the meridional plane inside the black hole.}
\label{theta-plusminus}
\end{figure}

Second, for functions which cannot be extended holomorphically below the horizon, one has to cover the interior region by suitable coordinates, write the relevant equations in them and solve the latter for the required functions. In our problem, this applies mainly to the second metric function $\lambda$ (and thus to various scalars that contain it). In the previous paper \cite{BasovnikS-16}, it proved useful to cover the meridional plane, inside the given black hole, by angular coordinates $\theta_-$ and $\theta_+$ ($0\!<\!\theta_-\!<\!\theta_+\!<\!\pi$), according to the transformation
\begin{align*}
  &r=M\left(1+\cos\frac{\theta_+-\theta_-}{2}\right),
   \quad
   \theta=\frac{\theta_++\theta_-}{2} \;, \\
  &{\rm i}\rho=\frac{M}{2}\,(\cos\theta_+-\cos\theta_-) \,,
   \quad
   z=\frac{M}{2}\,(\cos\theta_++\cos\theta_-) \,.
\end{align*}
In terms of these angles, the interior metric reads (it is no longer diagonal)
\begin{align}
  {\rm d}s^2=&-\left(1-\frac{2M}{r}\right)e^{2\nu_{\rm ext}}{\rm d}t^2  \nonumber \\
             &+r^2 e^{-2\nu_{\rm ext}}
               (e^{2\lambda_{\rm ext}}{\rm d}\theta_+{\rm d}\theta_-+\sin^2\theta\,{\rm d}\phi^2)\,,
  \label{metric,theta12}
\end{align}
where $r\!=\!r(\theta_+,\theta_-)$.
Geometrical meaning of the new coordinates is tied to the null geodesics which start tangentially to the horizon and inspiral towards the singularity, namely $\theta_+$ and $\theta_-$ are the latitudes on the horizon from where the geodesics intersecting at a given position $(r,\theta)$ start -- see figure 1 in \cite{BasovnikS-16} and figure \ref{theta-plusminus} here.

In the new coordinates, the equation for $\lambda_{\rm ext}\!:=\!\lambda\!-\!\lambda_{\rm Schw}$ (where $\lambda_{\rm Schw}$ is the value due to the given black hole alone) reads (see equation (57) in \cite{BasovnikS-16})
\begin{equation}
  \frac{\partial^2\lambda_{\rm ext}}{\partial\theta_+\partial\theta_-}=
  \frac{M(\nu_{{\rm ext},\theta_+}-\nu_{{\rm ext},\theta_-})}{2\,\sqrt{r(2M-r)}}
  -\nu_{{\rm ext},\theta_+}\nu_{{\rm ext},\theta_-}
\end{equation}
(this is actually an integrability condition for the gradient of $\lambda$ provided by Einstein equations).

Finally, one should not forget that the black-hole interior is a {\em dynamical} region, so it is not as obvious which section to portray (as obvious as in the external region where $t\!=\!{\rm const}$ is a clear choice because of $t$ being Killing time). In order to cover all the interior down to the singularity, one cannot manage with any cut which would everywhere be space-like (like in the exterior). It thus seems natural to choose a cut which is everywhere time-like -- and the simplest of these is to keep $t\!=\!{\rm const}$ like outside.

\begin{figure*}
\includegraphics[width=0.83\textwidth]{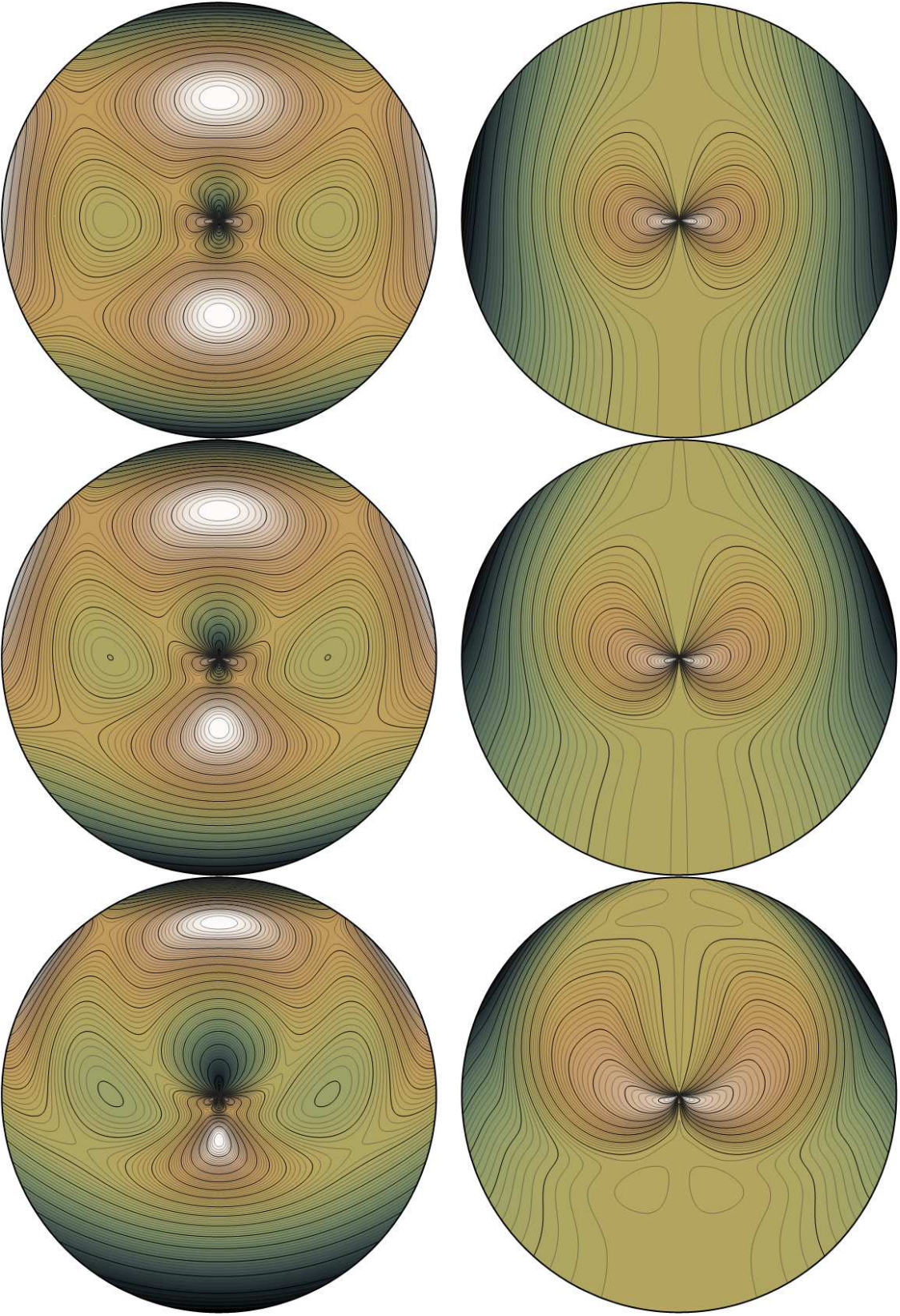}
\caption
{Meridional $t\!=\!{\rm const}$ contours of $\nu$ and $\lambda$ (left and right column) inside the ``bottom'' black hole, omitting the (divergent) parts $\nu_{\rm Schw}(z\!+\!h)$, $\lambda_{\rm Schw}(z\!+\!h)$ {\em due to that black hole alone}. Colouring is ``geographical'', positive values everywhere. In the top \!/\! middle \!/\! bottom rows, we chose $({\cal M},h,M)$ $=$ $(1,0.8713,0.3)$ \!/\! $(9.053,2.1,2)$ \!/\! $(8.379,1.25,0.8333)$ in the units of $a$.}
\label{nuext-lambdaext}
\end{figure*}

\begin{figure*}
\includegraphics[width=0.84\textwidth]{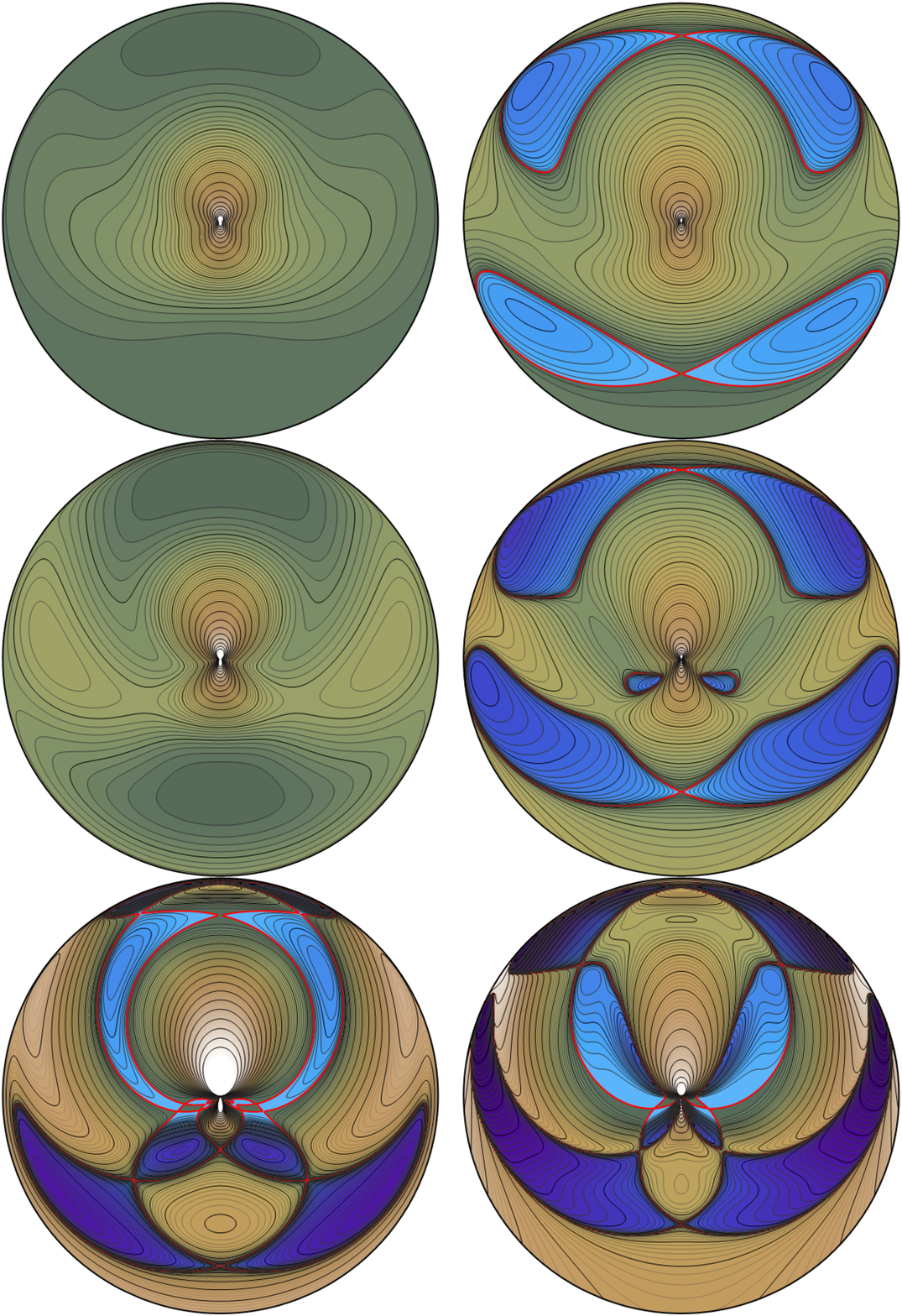}
\caption
{Meridional $t\!=\!{\rm const}$ contours of the acceleration and Kretschmann scalars $\kappa^2$ and $K$ (left/right column) inside the ``bottom'' black hole. Geographical colouring, with positive and negative values separated by a red border. Top \!/\! middle \!/\! bottom rows correspond to parameters $({\cal M},h,M)$ $=$ $(3,1.1515,0.9)$ \!/\! $(9.053,2.1,2)$ \!/\! $(8.379,1.25,0.8333)$ in the units of $a$.}
\label{acc-Kretschmann}
\end{figure*}

\subsection{Lapse function}

The lapse function of the total field is $N\!=\!e^\nu$, with the potential $\nu$ given by superposition (\ref{superposition}) (hence, the total lapse is given by product of the lapse functions ``generated" by individual sources). It is somewhat difficult to plot the total-potential (or lapse) contours, because its pure-Schwarzschild part behaves too wildly in comparison with the external contribution. We thus rather plot, in figure \ref{nuext-lambdaext}, just the external part, i.e. that due to the Appell ring and the other black hole. We depict the interior of the ``bottom'' black hole, so the Appell ring and the symmetric black hole are (would be) at the top in the figure. The left column shows $\nu_{\rm ext}$ and the right column shows $\lambda_{\rm ext}$, choosing three different parameter values (the three rows) -- see figure caption. Let us remind that since $\nu$ superposes linearly, $\nu_{\rm ext}$ is really just the part due to ``other'' sources, whereas $\lambda$ does not superpose linearly, so $\lambda_{\rm ext}$ also contains a non-linear, interaction part. We use ``geographical'' colouring, going -- in principle -- from brown (heights) to blue (depths).

\subsection{Gravitational acceleration}

Below the horizon, when using the $(t,\theta_+,\theta_-,\phi)$ coordinates, relation (\ref{acceleration}) reads
\begin{align}
  \kappa^2 &= 2g^{\theta_+\theta_-}N_{,\theta_+}N_{,\theta_-} = \nonumber \\
           &= \frac{e^{4\nu_{\rm ext}}}{r^4 e^{2\lambda_{\rm ext}}}\,
              (M\!+2\sqrt{\mbox{--}\Delta}\,\nu_{{\rm ext},\theta_+})
              (M\!-2\sqrt{\mbox{--}\Delta}\,\nu_{{\rm ext},\theta_-}) \,,
\end{align}
where $\Delta\!:=\!r^2\!-\!2Mr$ and, so,
\[\sqrt{\mbox{--}\Delta}=\sqrt{r(2M-r)}=M\,\sin\frac{\theta_+-\theta_-}{2} \;.\]

\subsection{Curvature}

In vacuum space-times the Ricci tensor vanishes and, especially in the static case, there is only one non-trivial quadratic curvature invariant -- the Kretschmann scalar $K\!:=\!R_{\mu\nu\kappa\lambda}R^{\mu\nu\kappa\lambda}$. We gave its various expressions in previous papers, just repeating one of them in (\ref{Kretschmann-static-Weyl}). Below the horizon, we again transform to the $(t,\theta_+,\theta_-,\phi)$ coordinates and obtain
\begin{equation}
  K=12\,({R^{\theta_-\theta_+}}_{\theta_-\theta_+})^2+
    16\,{R^{t\theta_-}}_{t\theta_+}{R^{t\theta_+}}_{t\theta_-} \;.
\end{equation}

The $t\!=\!{\rm const}$ meridional sections of the acceleration (actually its square, $\kappa^2$) and Kretschmann-scalar ($K$) contours are shown in figure \ref{acc-Kretschmann}, the left column containing $\kappa^2$ and the right column containing $K$, while the rows represent three different choices of the parameters (which correspond to gradual increase of the strain the black hole is subjected to). The whole range of geographical colouring is employed here, since both the acceleration square and Kretschmann scalar get negative in certain regions. In the former case it simply means that the gradient of $N$ (or $\nu$) is time-like in those regions. Negative values of the Kretschmann scalar are not very usual, but they have already been met in the literature. We also had this experience in previous paper \cite{BasovnikS-16} and tried to interpret it there by analysing how the individual components of the Riemann tensor contribute. In particular, we learned that it is {\em electric-type} components what makes the result negative, rather than the magnetic-type ones. Here we are arriving at similar conclusion: there again occur negative-$K$ zones inside the black hole, although there is no rotation in space-time. However, we again found such regions to {\em only} occur inside the black hole, which seems to indicate that {\em some} type of ``dragging'' -- here connected with dynamical nature of the black-hole interior -- has to be present (cf. \cite{CherubiniBCR-02}).

Let us also remind that $\kappa$ is uniform over the horizon (so the horizon is actually also a contour) and that the regions of negative $K$ touch the horizon at points where the Gauss curvature of the $t\!=\!{\rm const}$ section of the horizon vanishes (see the preceding paper \cite{BasovnikS-16}).

\section{Concluding remarks}

After considering, in previous two papers, an extreme black hole within the Majumdar-Papapetrou binary and a Schwarzschild-type black hole encircled by a Bach--Weyl thin ring, we have now subjected a black hole to a strain providing a static equilibrium to a system of two (or actually four) black holes kept in their positions by a ``repulsive'' effect of an Appell thin ring. We first confirmed that such a system can rest in a strut-free configuration and then studied its various properties. Focusing mainly on the geometry of black-hole interior (specifically, geometry of its $t\!=\!{\rm const}$ sections), we employed the same method as in the preceding paper \cite{BasovnikS-16}, namely integration of the relevant Einstein equations along special null geodesics which connect the horizon with the singularity.

The geometry within this system of sources is quite strongly deformed, as illustrated in figures showing meridional-plane contours 
of the lapse/potential, acceleration and Kretschmann scalars. In particular, if the system is sufficiently ``dense'' (meaning that the sources are close to each other and have sufficient masses), the Kretschmann scalar turns negative in some regions inside the black holes. Such a circumstance have already been met in the preceding paper, and there we also interpret it in terms of the nature of the relevant Riemann-tensor components and using the relation between the Kretschmann scalar and the Gauss curvature of the horizon.

Besides the option to subject a black hole to a yet another strong source of gravity, two apparent plans arise: to check whether a ``negative'' system made of {\em two} Appell rings placed symmetrically with respect to a black hole can also be in a strut-free equilibrium (and compare its properties with those of the present configuration), and to try to understand the regions of negative Kretschmann scalar on a more generic and fundamental (geometrical) manner.

\begin{acknowledgments}
We are grateful for support from the Czech grants GACR-14-37086G (O.S.); GAUK-369015 and SVV-260320 (M.B.).
O.S. also thanks T. Ledvinka for advice concerning the {\sc MAPLE} program and D. Bini for hospitality at Istituto per le Applicazioni del Calcolo ``Mauro Picone", CNR Roma.
\end{acknowledgments}

\bibliography{deformed-BHs-3.bib}

\end{document}